# Diagrammatic method for theory of magnetic and resistive properties of manganites.

E.E. Zubov[1,*], V.P. Dyakonov[1,2]

[1] *A.A. Galkin Donetsk Physics&Technology Institute NASU, 83114 Donetsk, Ukraine*

[2] *Institute of Physics, PAS , 02- 669 Warsaw , Poland.*

**Abstract.**

Effective field theory of magnetic and resistive properties of manganites with account of strong Hund exchange coupling and electron-phonon interactions has been evolved under the strong Hund coupling condition. In parallel with Lang-Firsov unitary transformation of the zeroth Hamiltonian, we have realized the diagonalization of Hund's Hamiltonian neglecting the upper triplet. The diagram techniques taking into account the quantum spin fluctuations of lower quintet and hole state with spin S = 3/2 was developed. The magnetic structure of the ground state and an influence of electron-phonon interaction have been analyzed using the first nonvanishing approximation of perturbation theory. Since a simple self-consistent equation for the Green's function is lacking, two approximations for effective interaction line were used, one of which is based on the assumption of Green's function symmetry with respect to change of polaron energy sign near the phase transition. The calculated temperature dependence of resistivity agrees well with experimental data including these obtained in applied magnetic field.

---

[*] e-mail: zubov@dyakon.fti.ac.donetsk.ua

## 1. Introduction.

At present the investigation of thermodynamics of the double-exchange model (DE) is an actual problem because the theory taking into account sequentially the quantum nature of electron and ion spins is still lacking. In manganites with dominant DE and strong electron correlations, strong Hund`s rule coupling of collective $e_g$ electron with the $Mn^{4+}$ ion spin should be considered. This essentially extends the wave function basis used and creates some difficulties in construction of diagram techniques. Anderson and Hasegawa[1] have calculated exactly the spectrum of electron excitation in system of two multivalent ions $Mn^{3+}$ and $Mn^{4+}$. De Gennes[2] has studied the thermodynamics of this system for a system of classical spins neglecting strong electron correlations. Kubo and Ohata[3] have proposed an exact projective transformation of Hamiltonian with Hund's rule coupling. Unfortunately, because of complicated dependence of this transformation on charge and spin degrees of freedom the description of thermodynamics was possible only in low temperature approximation. In the dynamic mean field approximation (DMF), the spins are as a rule classical [4]. Moreover, the account of kinematic electron contribution presupposes both an infinite space dimension and Dyson's method of diagram summation. In the coherent potential approximation (CPA) [5,6], the nonzero value of imaginary part of the Green's function on the Fermi level was obtained, and the sum rule for spectral density did not hold. One can suppose that evidently the contribution of charge and spin fluctuations was overestimated.

In this paper, the Hamiltonian investigated includes the strong Hund's rule coupling of localized $t_{2g}$ electrons of $Mn^{3+}$ ion with $e_g$ electrons, superexchange of localized spins, electron-phonon interactions as well as the contributions from phonon subsystem and interaction with applied magnetic field, h. In the diagrammatic method used the effective self-consistent field is extracted.

The thermodynamics of the DE model with account of Hund's rule coupling in the mean field approximation was previously considered in papers[7-9]. All contributions to the total Green's functions in the first nonvanishing approximation with respect to the inverse effective radius of interaction r ~ 1/z, where z is the number of nearest neighbour in the simple cubic (s.c.) lattice, were determined.

It should be emphasized that up to now the quantum fluctuations in the effective field theory as applied to manganites were not taken into account. We have calculated all quantum states corresponding to lower quintet of zeroth Hund's Hamiltonian neglecting the influence of the upper triplet. It is valid in the limit of infinite Hund's rule coupling $J_H$.

As a basis for the zeroth Hamiltonian, Hund's exchange and its additive part involved with chemical potential were used. In this theory, the Hamiltonian V describing the kinetic electron energy is a perturbation. We suppose that in the system studied the strong electron correlations are realized, and then one can neglect the states with twofold site filling. In a weakly doped electron subsystem with electron concentration n ~1, the Fermi level lies near the top of valence band. Then the chemical potential μ is proportional to W, where W is the bandwidth. In this case, the zeroth Hamiltonian, $H_0$, meets the necessary criteria of the perturbation theory, $H_0 >> V$.

## 2. The Hamiltonian of the system.

In the DE model with electron-phonon interactions the Hamiltonian of the system takes the form :

$$\hat{H} = \hat{H}_f + \hat{H}_b \ , \tag{1}$$

where the Fermi part, $\hat{H}_f$, is expressed as follows :

$$\hat{H}_f = -\sum_i J_H \boldsymbol{S}_i \boldsymbol{\sigma}_i - \sum_{i,j} J_{ij} (\mathbf{S}_i + \boldsymbol{\sigma}_i)(\mathbf{S}_j + \boldsymbol{\sigma}_j) - h \sum_i \left(\mathbf{S}_i^z + \boldsymbol{\sigma}_i^z\right) - \mu \sum_i n_i + V \tag{2}$$

The Bohr magneton, $\mu_B$, and $\tilde{g}$-factor were taken in unit system with $\mu_B \tilde{g} = 1$. In Hamiltonian (2) the second term includes the indirect exchange interaction of $e_g$ and $t_g$ electrons. Indeed, for $e_g$ and $t_g$ electrons we can consider the strong one-site Coulomb interaction U>>W and kinetic energy of intersite motion. Then it is easy to obtain the Heisenberg Hamiltonian in the form similar to the second term of Hamiltonian (2). To have correct limit as n→1 corresponding to pure $LaMnO_3$ the indirect exchange must also be expressed as in (2).

The perturbation Hamiltonian may be written as

$$\hat{V} = \sum_{i,j,\sigma} t_{ij} c_{\sigma i}^{+} c_{\sigma j}, \tag{3}$$

where $c_{\sigma i}^{+}$ $(c_{\sigma i})$ creates (annihilates) an electron of spin σ on lattice site i. The boson part of Hamiltonian (1) has a form similar to that used in the theory of small polaron:

$$\hat{H}_b = -g \sum_i n_i \left(b_i^{+} + b_i\right) + \omega_0 \sum_i b_i^{+} b_i, \tag{4}$$

where g is the electron-phonon coupling strength. In the Einstein model the phonon frequency $\omega_0$ is assumed to be dispersion-free. $b_i^+$ и $b_i$ are the phonon creation and annihilation operators.

In the theory proposed, the following relation of parameters : $J_H >> t_{ij} \sim g >> |J_{ij}| \sim h$ was used. At the first glance the operator V is not the perturbation. However, as will be seen in the further consideration the hopping integral $t_{ij}$ enters into the expression for chemical potential μ which is proportional to bandwidth W = 2zt, where t is the nearest-neighbour hopping integral. The zeroth Hamiltonian contains only a part of V, which does not depend on the free carriers concentration. The perturbation near the band half filling is proportional to hole concentration, since the electron jumps at n = 1 are forbidden for this strongly correlated system. The foregoing provides the basis for the construction of perturbation theory at n ~ 1, with the results correct to within 1/z.

The electron-phonon interactions in manganites were first taken into account in article[10] assuming the phonon to be a localized classical oscillator. The main conclusion obtained is that in $La_{1-x}Sr_xMnO_3$ the pure DE model is usable, while in $La_{1-x}Ca_xMnO_3$ the strong electron-phonon interaction plays an essential role.

In the following, the part of Hamiltonian (2) connected with superexchange interaction is considered in the mean field approximation. We will carry out the unitary transformation $\tilde{U} = \exp\left(\tilde{S}\right)$ of Hamiltonian (1) resulting in the fermion and boson operators separation from each other. The expression for $\tilde{S}$ given by Lang and Firsov [11] has the following form:

$$\tilde{S} = -\frac{gn}{\omega_0}\sum_i (b_i^+ - b_i) \qquad (5)$$

The shift of Bose

$$\tilde{b} = b + \frac{gn}{\omega_0} \qquad (6)$$

and multiplying Fermi operators

$$\tilde{c}_\sigma = Yc_\sigma \ , \qquad (7)$$

where $Y = e^{\lambda(b^+ - b)}$ connected exceptionally with the phonon degree of freedom, $\lambda = g/\omega_0$, results from the transformation $\tilde{S}$. The Hermitian conjugate of Eq. (7) is an expression for creation operator. Substituting the transformed operators in Eq. (1) we obtain the following zeroth Hamiltonian for fermion subsystem:

$$\begin{aligned}\hat{\tilde{H}}_{0f} = \sum_i J_H \mathbf{S}_i \boldsymbol{\sigma}_i - 2J(0)\sum_{ij}(\mathbf{S}_i + \boldsymbol{\sigma}_i)(< S_i^z > + < \sigma_i^z >) - \\ - \mu\sum_i n_i - h\sum_i (S_i^z + \sigma_i^z) - \xi\sum_i n_i^2\end{aligned} \ , \qquad (8)$$

where J(0) = zJ in the nearest-neighbour approximation, $\xi = g^2/\omega_0$ is the polaron binding energy, $n_i$ the number of $e_g$ electrons on the i-th site. The boson part has the form :

$$\hat{H}_{0b} = \omega_0 \sum_i b_i^+ b_i \tag{9}$$

The Hamiltonian of interaction V as a function of $c_\sigma$ и $c_\sigma^+$ is expressed in terms of operators $\tilde{c}_\sigma$ и $\tilde{c}_\sigma^+$, respectively.

We will consider the transformed Hamiltonian (1) where the sign of tilde was omitted. Preliminarily we should carry out a diagonalization of Hamiltonian (8) for the Fermi subsystem. The main difficulty is associated with the first term describing Hund's rule coupling between the ion core and mobile electron. The total basis of the zeroth Fermi Hamiltonian (8) includes 12 spin wave functions supposing that $Mn^{4+}$ ion spin S = 3/2 and $e_g$ electron spin σ = 1/2. At h = 0, in the Hund's part of Eq. (8) the eight spin functions correspond to five- and three-fold degenerated levels of $E_H^0 = -\frac{1}{2} S J_H$ and $E_H^2 = \frac{1}{2}(S+1) J_H$ with $S_2 = 2$ and $S_1 = 1$ spins, respectively. Spin S = 3/2 corresponds to hole state ($Mn^{4+}$ ion) with four-fold degenerated energy level of $E_H^1 = 0$. The wave functions for the above multiplets are[12]:

$$\left|(3/2,1/2)S'm\right\rangle = \left|\varphi_{S'm}\right\rangle = \sum_{m_1 m_2} C_{m_1\ m_2\ m}^{3/2\,1/2\,S'} \left|2m_1, 2m_2\right\rangle, \tag{10}$$

where $\left|2m_1, 2m_2\right\rangle = \left|3/2, m_1\right\rangle \otimes \left|1/2, m_2\right\rangle$, $C_{m_1\ m_2\ m}^{3/2\,1/2\,S'}$ are the Clebsch-Gordan coefficients. In Eq.(10), at first, $m_2$ and then $m_1$ are changing from maximum to minimum values. Setting such a numeration order the vectors $\left|2m_1, 2m_2\right\rangle$ can convert to $\left|\psi_i\right\rangle$, where index i=1, 2…12. The first five vectors correspond to quintet $E_H^0$, the next three vectors belong to triplet $E_H^2$ with electron spin being antiparallel to ion spin and the last four vectors describe the hole state $E_H^1$ with spin S = 3/2.

Let us introduce the Hubbard's operators $X^{ik} = |\psi_i\rangle\langle\psi_k|$. Then the matrix $\hat{\mathbf{C}}$ of Clebsch-Gordan coefficients in Eq.(10) may be presented as

$$\hat{\mathbf{C}} = X^{1,1} + X^{5,8} + X^{9,9} + X^{10,10} + X^{11,11} + X^{12,12} + \frac{1}{2}(X^{2,2} + X^{4,7} - X^{6,3} + X^{8,6}) + \frac{\sqrt{3}}{2}(X^{2,3} + X^{4,6} + X^{6,2} - X^{8,7}) + \frac{1}{\sqrt{2}}(X^{3,4} + X^{3,5} + X^{7,4} - X^{7,5}) \quad (11)$$

The vector system can be written as $|\psi_i\rangle$: $|3,1\rangle, |3,-1\rangle, |1,1\rangle, |1,-1\rangle, |-1,1\rangle, |-1,-1\rangle, |-3,1\rangle, |-3,-1\rangle$, where i = 1,2…8, for electron states, and $|\psi_i\rangle$: $|3,0\rangle, |1,0\rangle, |-1,0\rangle, |-3,0\rangle$, where i = 9,10,11,12 for hole states. The vectors $|\psi_i\rangle$ are connected with eigenfunctions $|\varphi_i\rangle$ of the Hund's part in Hamiltonian (8) by linear relation

$$|\varphi_k\rangle = \sum_i C_{ki} |\psi_i\rangle \ , \quad (12)$$

where coefficient $C_{ki}$ forms a matrix $\hat{\mathbf{C}}$ in Eq. (11).

Using the function $|\varphi_k\rangle$ the Hamiltonians (2) and (3) will be transformed. In Eq. (2), superexchange interaction includes z-projection of total spin operator. Then in $|\varphi_k\rangle$ basis the latter will be presented as a direct sum of diagonal operators $S^z \oplus S_1^z \oplus S_2^z$ for spins S = 3/2, $S_1$ = 1 и $S_2$ = 2 [12], respectively. Having constructed system of Hubbard's operators $L^{ik} = |\varphi_i\rangle\langle\varphi_k|$, one can write the unitary transformed zeroth Hamiltonian in the diagonal form:

$$\hat{\tilde{H}}_0 = \sum_{i=1}^{N} \left\{ \sum_{l=1}^{5} \varepsilon_l L_i^{ll} + \sum_{l=9}^{12} \varepsilon_l L_i^{ll} \right\}, \quad (13)$$

where $\tilde{H} = h + 2J(0)(<S^z> + <\sigma^z>)$ is the sum of applied magnetic and effective Weiss fields, $\varepsilon_l = -\frac{1}{2}SJ_H - \mu - \xi - (2S - l)\tilde{H}$ for state with spin $S_2$ = 2 ($l \leq 5$) and $\varepsilon_l = -(S + 9 - l)\tilde{H}$ for hole state with S = 3/2 ($l \geq 9$). At the same time we have taken into account that $J_H >> t >> J$. Therefore the triplet level with $S_1$ = 1 lies considerably above the state with $S_2$ = 2 and can be neglected. We also ignore the contribution of chemical potential $-\mu(L^{6,6} + L^{7,7} + L^{8,8})$. The above statements can be more strictly proved with the introduction of projective operator

$$P = \sum_{\alpha=1}^{5} L^{\alpha\alpha} + \sum_{\alpha=9}^{12} L^{\alpha\alpha} \tag{14}$$

Acting by the P operator on both the Hamiltonian (1) and wave functions to an accuracy of $t / J_H$ we eliminate all states with numbers $\alpha$ = 6,7 и 8.

The operator $\mathbf{A}$ in the matrix form can be presented as

$$\mathbf{A} = \sum_{lm} \langle l | A | m \rangle X^{lm}$$

One can write the following expressions for electron creation operators:

$$\begin{aligned} c_\uparrow^+ &= X^{1,9} + X^{3,10} + X^{5,11} + X^{7,12} \\ c_\downarrow^+ &= X^{2,9} + X^{4,10} + X^{6,11} + X^{8,12} \end{aligned} \tag{15}$$

The corresponding formulae for electron annihilation operators can be obtained using the Hermitian conjugate of Eq. (15). The Hubbard's operators $X^{ik}$ и $L^{lm}$ are related to each other by unitary transformation:

$$X^{ik} = \sum_{lm} C_{li}^* C_{mk} L^{lm} \ , \tag{16}$$

where coefficients are determined by matrix $\hat{\mathbf{C}}$ from Eq. (11). Substituting (16) in (15) we find the expressions for unitary transformed $\tilde{c}_\uparrow^+$ and $\tilde{c}_\uparrow$ operators as functions of $L^{lm}$. An explicit form of both $c_\uparrow^+$ and $c_\uparrow$ operators and electron and ion spin operators is given in Appendix. The operator of electron number, n, is invariant relative to unitary transformation (12).

## 3. Fermion-boson free-particle Green's function and effective kinematic interaction.

In investigation of electron dynamics we have used the Matsubara Green's functions:

$$\mathcal{G}_{\sigma}(\tau, r_l - r_m) = -< T_\tau c_{l\sigma}(\tau) c^+_{m\sigma}(0) >, \qquad (17)$$

where $T_\tau$ is the chronological ordering operator. The operators $c^+_\uparrow$ and $c_\uparrow$ are expressed in the Heisenberg representation. The brackets <…> mean that the Gibbs thermodynamic average is obtained using the total Hamiltonian (1).

The problem of finding of Green's functions reduces to the calculation of various correlators appearing in the scattering matrix series expansion of perturbation theory with Hamiltonian V in interaction representation. Wick's theorem[13,14] for Hubbard's operators is used for the unlinking of correlators, and then the task reduces to the calculation of elementary Green's functions. Every contribution of the perturbation theory has its graphical form. In accordance with Eq. (7) the free-particle Green's functions $U(\tau, \varepsilon_{lm})$ are determined as follows :

$$U(\tau, \varepsilon_{lm}) = -< T_\tau Y(\tau) Y(0) >_{0ph} < T_\tau L^{l,m}(\tau) L^{m,l}(0) >_0 \frac{1}{< F^{l,m} >_0}, \qquad (18)$$

where $< F^{l,m} >_0 = < L^{l,l} + L^{m,m} >_0$, $\varepsilon_{lm} = \varepsilon_l - \varepsilon_m$. The thermal averages of the first and second correlators in Eq. (18) are calculated with Hamiltonians (9) and (13), respectively. The Fermi part of Eq. (18) is easily derived using Wick's theorem :

$$G^{lm}_{0el.}(\tau) = < T_\tau L^{l,m}(\tau) L^{m,l}(0) >_0 \frac{(-1)}{< F^{l,m} >_0} = e^{\varepsilon_{lm}\tau} \cdot \begin{cases} -f(\varepsilon_{lm}), & \tau > 0 \\ 1 - f(\varepsilon_{lm}), & \tau < 0 \end{cases},$$

where $f(x) = 1/(e^{\beta x} + 1)$ is the Fermi distribution function, $1/\beta = T$ the temperature.

The Bose part of Eq. (18) may be determined using both the relation for boson operators b+ и b: $e^{\lambda(b^+ - b)} = e^{-\frac{1}{2}\lambda^2} e^{\lambda b^+} e^{-\lambda b}$ and the Feynman disentangling of these operator products. The detailed calculation is given in Ref.[15]

As is seen in Eq. (18), the Bose part has the form of $G_0^{ph.}(\tau) = < T_\tau Y(\tau) Y(0) >_{0ph} = e^{\Phi(\tau)}$ , where

$$\Phi(\tau) = -\lambda^2 \left\{ 2B + 1 - 2\sqrt{B(B+1)} \cosh\left[ \omega_0 \left( \tau \mp \frac{\beta}{2} \right) \right] \right\}$$

and the upper minus and the lower plus signs correspond to $\tau > 0$ and $\tau < 0$, respectively. $B = 1/(e^{\beta\omega_0} - 1)$ is the Bose distribution function for Einstein phonon mode. The appearance of hyperbolic cosine instead of ordinary, as in Ref.15, is because in Matsubara formalism the time t is imaginary, and therefore we must make the substitution *t→ it.* It is interesting to point out that the function $G_0^{ph.}(\tau)$ describes nondiagonal transitions by Holstein definition[16], which are responsible for number of phonons changing in the hopping process. In the function $G_0^{ph.}(\tau)$, the time-independent Debye-Waller factor $e^{-\lambda^2(2B+1)}$ corresponds to diagonal transitions with no changes in the number of virtual phonons.

The Fourier transformation of single-particle Green's function has the form $U(i\omega_n, \varepsilon_{lm}) = \frac{1}{2\beta} \int_{-\beta}^{\beta} U(\tau, \varepsilon_{lm}) e^{i\omega_n \tau} d\tau$ . It is easy to write the following relation

$$e^{x \cosh(z)} = \sum_{k=-\infty}^{\infty} I_k(x) e^{kz} ,$$

where $I_k(x)$ are the Bessel functions of complex argument. Note that in an initial time space we have a simple product of single-particle Green's functions of Fermi and Bose subsystems. In Fourier inverse space this relation is complicated:

$$U(i\omega_n,\varepsilon_{lm})=\frac{1}{\beta}f(\varepsilon_{lm})e^{-\lambda^2(2B+1)}\sum_{k=-\infty}^{+\infty}I_k\left(2\lambda^2\sqrt{B(B+1)}\right)\frac{e^{\beta\varepsilon_{lm}+\frac{1}{2}\beta k\omega_0}+e^{-\frac{1}{2}\beta k\omega_0}}{i\omega_n+k\omega_0+\varepsilon_{lm}}, \tag{19}$$

where $\omega_n=\pi(2n+1)\beta$ and unit system was taken for $k_B$ = $\hbar$ = $1$. Since the operators $c_\sigma^\pm$ and $L^{l,m}$ are linearly connected (see Eqs. (A.1) in Appendix) one can write the single- particle Green's function as a linear combination of functions $U(i\omega_n,\varepsilon_{lm})$:

$$\tilde{G}_{0\sigma}(i\omega_n)=-\left\{\begin{matrix}(\delta_{j,0}+\frac{\delta_{j,1}}{4})<F^{9,1+j}>_0U(i\omega_n,\varepsilon_{9,1+j})+\frac{3-j}{4}<F^{10,2+j}>_0U(i\omega_n,\varepsilon_{10,2+j})+\\+\frac{2+j}{4}<F^{11,3+j}>_0U(i\omega_n,\varepsilon_{11,3+j})+\left(\frac{\delta_{j,0}}{4}+\delta_{j,1}\right)<F^{12,4+j}>_0U(i\omega_n,\varepsilon_{12,4+j})\end{matrix}\right\},$$

(20)

where index j = 0 for σ = +1 (spin up) and j = 1 for σ = -1 (spin down) and $\delta_{i,j}$ is the Kronecker symbol. The following identities: $\varepsilon_{9,1}=\varepsilon_{10,2}=\varepsilon_{11,3}=\varepsilon_{12,4}=\tilde{\mu}+\frac{1}{2}\tilde{H}=\varepsilon_{0\uparrow}$ and $\varepsilon_{9,2}$=$\varepsilon_{10,3}$=$\varepsilon_{11,4}$=$\varepsilon_{12,5}$= $\tilde{\mu}-\frac{1}{2}\tilde{H}=\varepsilon_{0\downarrow}$ are valid, where $\tilde{\mu}=\mu+\xi+\frac{1}{2}SJ_H$. Here the change of electron energy with spin σ transferring from site to site is denoted by $\varepsilon_{0\sigma}$. Then expression (20) is simplified and takes the form:

$$\tilde{G}_{0\sigma}(i\omega_n)=<F^{\sigma 0}>_0G_{0\sigma}(i\omega_n)=<F^{\sigma 0}>_0U(i\omega_n,\varepsilon_{0\sigma}), \tag{21}$$

where a combined occupancy of electron-hole states $<F^{\sigma 0}>_0$ in the mean field approximation is defined as

$$<F^{+0}(\varepsilon_1,\varepsilon_2,\varepsilon_3,\varepsilon_4,\varepsilon_5,\varepsilon_9,\varepsilon_{10},\varepsilon_{11},\varepsilon_{12})>_0=<F^{9,1}>_0+\frac{3}{4}<F^{10,2}>_0+\frac{1}{2}<F^{11,3}>_0+\frac{1}{4}<F^{12,4}>_0$$
$$<F^{-0}(\varepsilon_1,\varepsilon_2,\varepsilon_3,\varepsilon_4,\varepsilon_5,\varepsilon_9,\varepsilon_{10},\varepsilon_{11},\varepsilon_{12})>_0=\frac{1}{4}<F^{9,2}>_0+\frac{1}{2}<F^{10,3}>_0+\frac{3}{4}<F^{11,4}>_0+<F^{12,5}>_0$$

(22)

For the sake of convenience we have written $<F^{\sigma 0}>_0$ as a function of all energy parameters $\varepsilon_l$. It can be shown that expression (21) coincides with unperturbed Green's function in work [6]. The equality

$$F^{\sigma 0} = n_\sigma + p_\sigma \ , \qquad (23)$$

where $n_\sigma$ and $p_\sigma$ are the operators of number of electrons and holes with spin σ expressed in terms of $L^{lm}$ (see Appendix), will be further used. Using equalities of (A.2-A.4) in Appendix, Eq. (23) can be written as :

$$<F^{\sigma 0}> = \frac{1}{8}(5-n) + \sigma <\sigma^z + \frac{1}{4}S_0^z>, \qquad (24)$$

where $S_0^z = \frac{3}{2}(L^{9,9} - L^{12,12}) + \frac{1}{2}(L^{10,10} - L^{11,11})$ is a z-projection of spin in basis of $t_{2g}$ electrons as in $Mn^{4+}$ ion. It is easy to check that expectations of operators $<\sigma^z + \frac{1}{4}S_0^z>$ and $\frac{1}{4}<\sigma^z + S^z>$ are equal (see Appendix). Thus, a quarter of $Mn^{4+}$ ion spin fluctuates with electron spins. The reason is that $Mn^{4+}$ ion spins are coupled to spins of itinerant $e_g$–electron not only by Hund's exchange but by effective kinematic interaction as well. This point of view will be confirmed in the following discussion.

We will formulate some rules to find the contributions of series of perturbation theory in both Green's function and combined occupancies using graphic representation. In Figs. 1 and 2, the electron Green's functions $G_{0\sigma}(i\omega_n)$, $G_{+-}(i\omega_n) = U(i\omega_n, \varepsilon_{\uparrow\downarrow})$ and Fourier components of interaction

$$t(\mathbf{q}) = \sum_{ij} t_{ij} e^{-i\mathbf{q}(\mathbf{r}_i - \mathbf{r}_j)} = 2t(\cos(q_x a) + \cos(q_y a) + \cos(q_z a)) \ ,$$

where $\varepsilon_{\uparrow\downarrow} = \varepsilon_{0\downarrow} - \varepsilon_{0\uparrow}$ and $a$ is the constant of s.c. lattice, are presented by solid, dashed and wave lines, respectively

Using Wick's theorem one can write all possible pairings realized in the framework of this consideration. The $c_\sigma$ operators and new Bose operator $B^+ = \sigma^+ + \frac{1}{2} S_0^+$ are "active" operators where

$$S_0^+ = \frac{\sqrt{3}}{2}\left(L^{9,10} + L^{11,12}\right) + L^{10,11},$$

$$\sigma^+ = \frac{1}{2}\left(L^{1,2} + L^{4,5}\right) + \sqrt{\frac{3}{8}}\left(L^{2,3} + L^{3,4}\right)$$

The operators $B^-$ and $\sigma^-$ appear due to the Hermitian conjugate of operators $B^+$ and $\sigma^+$ ($B^- = (B^+)^+$, $\sigma^- = (\sigma^+)^+$), respectively. We have the following zero pairings:

$$(n_{\sigma i} + p_{\sigma i})\underset{\longleftarrow}{\;} c_{\sigma j} = c_{\sigma_1 i}\underset{\longleftarrow}{\;} c_{\sigma_2 j} = c^+_{\sigma_1 i}\underset{\longleftarrow}{\;} c^+_{\sigma_2 i} = B^\sigma_i \underset{\longleftarrow}{\;} c_{\bar\sigma j} = 0 \qquad (25)$$

Here i and j site symbols mean the time indices in the interaction representation. In pairings the arrows are directed from "active" to "passive" operators. The remaining nonzero pairings have the form:

$$(n_{\sigma i} + p_{\sigma i})\underset{\longleftarrow}{\;} c_{\bar\sigma j} = \frac{1}{4}\delta_{ij} G_{0\bar\sigma}(\tau_j - \tau_i) c_{\bar\sigma i}, \quad c^+_{\sigma i}\underset{\longleftarrow}{\;} c_{\bar\sigma j} = \delta_{ij} G_{0\bar\sigma}(\tau_j - \tau_i) B^\sigma_i$$

$$B^\sigma_i \underset{\longleftarrow}{\;} c_{\sigma j} = -\frac{1}{4}\delta_{ij} G_{0\sigma}(\tau_j - \tau_i) c_{\bar\sigma i}, \quad (\sigma^z_i + \frac{1}{4} S^z_{0i})\underset{\longleftarrow}{\;} c_{\sigma j} = -\frac{\sigma}{8}\delta_{ij} G_{0\sigma}(\tau_j - \tau_i) c_{\sigma i} \qquad (26)$$

$$(n_{\sigma i} + p_{\sigma i})\underset{\longleftarrow}{\;} B^+_j = \frac{\sigma}{4}\delta_{ij} G_{+-}(\tau_j - \tau_i) B^+_i, \quad B^-_i \underset{\longleftarrow}{\;} B^+_j = -\frac{1}{2}\delta_{ij} G_{+-}(\tau_j - \tau_i)(\sigma^z_i + \frac{1}{4} S^z_{0i})$$

The presented expressions are closed and allow to find the contribution of series of perturbation theory using only three single-particle Green's functions in spite of the existence of a large number of Hubbard operators $L^{ik}$. This result is important and essentially simplifies the construction of diagram techniques in DE model with a strong Hund's rule coupling.

Let us find an analytic expression for the effective kinematic interactions $\beta B^{c^+_\sigma c_\sigma}(\mathbf{q}, i\omega_n)$ displayed by bold line in Fig.1. This expression may be written as

$$\beta B^{c_\sigma^+ c_\sigma}(\mathbf{q}, i\omega_n) = \frac{\beta t(\mathbf{q})}{1 - \beta t(\mathbf{q}) G_{0\sigma}(i\omega_n) < F^{\sigma 0} >} \qquad (27)$$

In the absence of electron-phonon interaction the calculation of diagrams presents no problems [8]. In this case, Eq. (27) has the simplest form

$$\beta B^{c_\sigma^+ c_\sigma}(\mathbf{q}, i\omega_n) = \frac{\beta t(\mathbf{q})(i\omega_n + \varepsilon_{0\sigma})}{i\omega_n - E_{\sigma\mathbf{q}}} \quad , \qquad (28)$$

where $E_{\sigma\mathbf{q}} = -\varepsilon_{0\sigma} + t(\mathbf{q}) < F^{\sigma 0} >$.

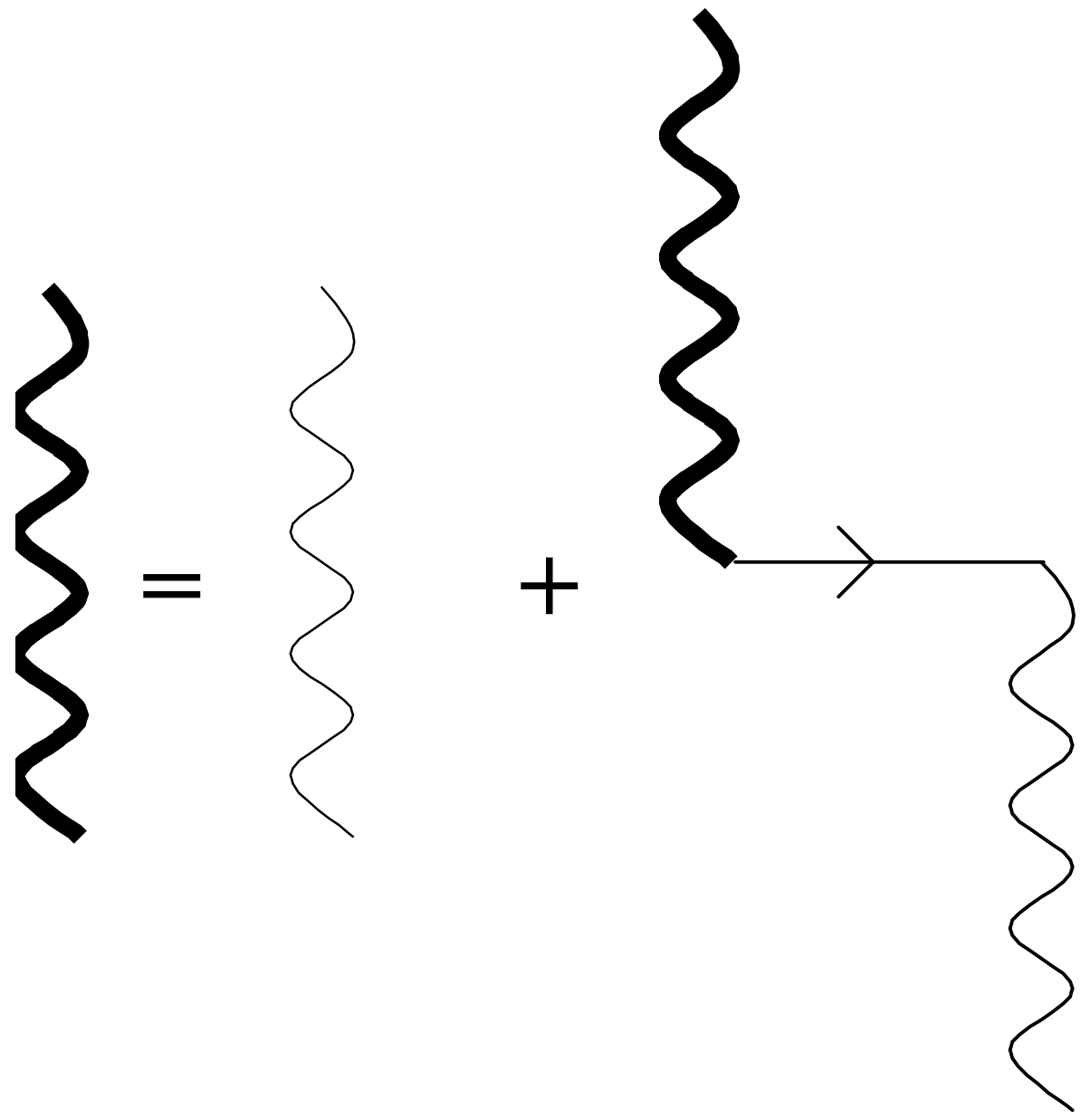

Fig.1. Graphic equation for effective kinematic interaction.

If Eq. (28) contains one pole, the infinite quantity of poles in Eq. (27) are the solutions of algebraic equation of infinite order. As a consequence, there is unlimited quantum number of phonons. With the aim of practical use Eq. (27) should be simplified.

## 4. Electron- phonon interaction.

In CPA method [6] the polaron problem is considerably simplified after analytic continuation $i\omega_n \to \omega + i\delta$ and use of self-consistent equation for the total Green's function $\mathcal{G}_\sigma(\omega,0) = U(\omega - \frac{W^2}{16}\mathcal{G}_\sigma(\omega,0),\varepsilon_{0\sigma})$. Writing $\mathcal{G}_\sigma(\omega,0)$ as a sum of real and imaginary parts it is easy to solve the derived set of equations by simple iteration method.

Let us consider two approximations in solving Eq. (27) for frequency poles. As will be seen later, the first approximation well works far from the phase FM-PM transition. However, in this case the calculated values of resistivity contradict the experimental data. It is not only because of an approximate solution of equation for electron-phonon effective line. In Kubo formula the two-particle correlator is decoupled in product of Green's functions $\mathcal{G}_\sigma(-\tau,\mathbf{r}_{j_2}-\mathbf{r}_{i_1})\mathcal{G}_\sigma(\tau,\mathbf{r}_{j_1}-\mathbf{r}_{i_2})$ (bubble approximation). Then the vertex corrections, which contribution in magnetoresistance effect may be essential, are not taken into account. This problem calls for further investigation and is not considered in this paper.

a) *First approximation*. The equation of poles singularities for the effective kinematic interaction (27) is written in the form:

$$\prod_m (x + m\omega_0) - t(\mathbf{q}) < F^{\sigma 0} > P_\sigma \sum_m \varphi_{m\sigma} \prod_{k(\neq m)} (x + k\omega_0) = 0, \qquad (29)$$

where

$$x = i\omega_n + \varepsilon_{0\sigma} \quad P_\sigma = f(\varepsilon_{0\sigma})e^{-\lambda^2(2B+1)}, \quad \varphi_{m\sigma} = I_m\left(2\lambda^2\sqrt{B(B+1)}\right)\left(e^{\beta\left[\varepsilon_{0\sigma}+\frac{1}{2}m\omega_0\right]} + e^{-\frac{1}{2}\beta m\omega_0}\right).$$

The second term in Eq. (29) is considered as perturbation. It is correct in the limit of strong electron-phonon coupling, when $P_\sigma\varphi_{m\sigma} << 1$. In the case of g = 0, using Eq. (29) we obtain the solution corresponding to DE model. In Eq. (29) the term with m = 0 corresponds to the central polaron band. Expressing the n-th root of Eq. (29) as $x = x_n + \Delta_n$, where $x_n = -n\omega_0$ is the root in the absence of perturbation, it is easy to calculate a correction to the n-th pole in linear approximation relative to smallness of $P_\sigma\varphi_{m\sigma}$ parameter

$$\Delta_n = t(\mathbf{q}) < F^{\sigma 0} > P_\sigma \varphi_{n\sigma} \tag{30}$$

Then one can write the equation (27) accurate to $t(q)^2$ as

$$\beta B^{c_\sigma^+ c_\sigma}(\mathbf{q}, i\omega_n) = \beta t(\mathbf{q}) \prod_m \frac{i\omega_n + \varepsilon_{0\sigma} + m\omega_0}{(i\omega_n - E_{m\sigma\mathbf{q}})} \approx \beta t(\mathbf{q}) \left\{ 1 + t(\mathbf{q}) < F^{\sigma 0} > P_\sigma \sum_{m=-\infty}^{\infty} \frac{\varphi_{m\sigma}}{i\omega_n - E_{m\sigma\mathbf{q}}} \right\}, \tag{31}$$

where $E_{m\sigma\mathbf{q}} = -\varepsilon_{0\sigma} - m\omega_0 + t(\mathbf{q}) < F^{\sigma 0} > P_\sigma \varphi_{m\sigma}$.

b) *Second approximation*. It is used to calculate the magnetoresistance effect near the phase transition. We suppose that constant $\lambda^2 >> 1$, i.e. electron-phonon coupling is sufficiently strong in comparison with phonon energy of vibration. The $U(i\omega_n, \varepsilon_{0\sigma})$ function can be presented in the form of infinite series expansion

$$U(i\omega_n, \varepsilon_{0\sigma}) = \frac{1}{\beta} e^{-\lambda^2(2B+1)} \sum_{k=0}^{\infty} \left( \frac{-2}{\beta} \right)^k \frac{\omega_0^k}{(i\omega_n + \varepsilon_{0\sigma})^{k+1}} \frac{d^k \psi}{d\omega_0^k} \eta_k, \tag{32}$$

where the derivative of zeroth-order $\frac{d^0\psi}{d\omega_0^0} = \psi = \exp\left\{ 2\lambda^2 \sqrt{B(B+1)} \cosh\left( \frac{\beta\omega_0}{2} \right) \right\}$, and $\eta_k = 1 - 2f(\varepsilon_{0\sigma})$ and $\eta_k = 1$ for odd and even k, respectively. Keeping the term with maximal power of $\lambda$ in expressions $\frac{d^k\psi}{d\omega_0^k}$, it is easy to obtain that

$$\frac{d^k\psi}{d\omega_0^k} \approx \psi\left[\frac{\beta}{2}\lambda^2\sqrt{B(B+1)}\sinh\left(\frac{\beta\omega_0}{2}\right)\right]^k = \psi\left(\frac{\beta\xi}{2}\right)^k \tag{33}$$

It should be noted that this approximation is valid at sufficiently low temperatures, when $\omega_0/T \sim 1$, since at high temperatures ($T >> \omega_0$) the Bose function has a value comparable to maximal power of parameter $\lambda$, and Eq. (33) will be wrong. Eq. (32) is exact if the inequality $\lambda^4 >> 2B+1$ is fulfilled. In this case the following relation $\frac{\omega_0}{T} >> \ln\frac{\lambda^4+1}{\lambda^4-1}$ is valid.

The high temperature range is investigated in detail using the small polaron theory [15,16]. For nondiagonal transition it was shown that a hopping rate has activation character. The value of gap is equal to one-half of the polaron binding energy. The most interesting temperature range is the $T \sim \omega_0$ region because the phase transition temperature is usually of the order of $\omega_0$.

By substituting Eq. (33) for Eq. (32) and replacing $\xi$ for $-\xi$ we obtain the terms of series with the opposite sign. Such asymmetry is quite typical for system with electron-phonon interactions [15] when polarons are formed. We will suppose that near transition in ordered state the processes of both formation and destruction of polarons would be apparently interchangeable. Therefore Green's function will be invariant relatively to the change of sign for polaron energy, $\xi$. Neglecting the terms with odd power of $\xi$ in Eq. (32), we obtain the ordinary geometric series which sum is easily calculated. The following expression for single particle Green's function with electron-phonon coupling was obtained :

$$U(i\omega_n, \varepsilon_{0\sigma}) = \frac{i\omega_n + \varepsilon_{0\sigma}}{\beta(i\omega_n + \varepsilon_{0\sigma} - \xi)(i\omega_n + \varepsilon_{0\sigma} + \xi)} \tag{34}$$

Eq. (34) has a simple physical meaning. At emission or absorption of phonon by ion coupled with itinerant electron, the chemical potential changes by value $\xi$. In this case the effective interaction line is determined as

$$\beta B^{c_\sigma^+ c_\sigma}(\mathbf{q}, i\omega_n) = \frac{\beta t(\mathbf{q})\left[(i\omega_n + \varepsilon_{0\sigma})^2 - \xi^2\right]}{(i\omega_n - E_{1\sigma\mathbf{q}})(i\omega_n - E_{2\sigma\mathbf{q}})}, \tag{35}$$

where $E_{1,2\sigma\mathbf{q}} = -\varepsilon_{0\sigma} + \frac{1}{2}\left\{t(\mathbf{q}) < F^{\sigma 0} > \pm\sqrt{t(\mathbf{q})^2 < F^{\sigma 0} >^2 + 4\xi^2}\right\}$. Using Eq. (35) one can calculate spectral density characteristic for Mott-Hubbard dielectric with the gap of the order of polaron energy $\xi$.

**5. The analysis of magnetic structure.**

Using the first and second approximations for the effective kinematic interaction, we will write the system of equations determining the chemical potential and mean spin of the system. The graphic image of series expansion for combined occupancies $<F^{\sigma 0}>$ is presented in Fig.. 2a. A small circle $\circ$ corresponds to $<F_i^{\sigma 0}>_0$ of the i-th site. The lower index for $<F_i^{\sigma 0}>_0$ denotes the averaging over Hamiltonian $\hat{\hat{H}}_0$ (Eq. (13)) with the parametric part tending to zero :

$$\hat{H}_0(r_\uparrow, r_\downarrow) = \hat{\hat{H}}_0 + \sum_i^N \left(r_{\uparrow i} F_i^{+0} + r_{\downarrow i} F_i^{-0}\right) \tag{36}$$

The dots connecting the diagram blocks (Fig. 2) mean the equality of their external site indices. According to the linked cluster theorem such diagrams characterize the contributions to both combined occupancies and Green's function. The form of zeroth Hamiltonian (36) is very suitable for calculation of operator average

$$< F_i^{\sigma 0} >_0 = \frac{\partial}{\partial(-\beta r_{\sigma i})} \ln Tr(\exp(-\beta \hat{H}_0(r_\uparrow, r_\downarrow)) = \partial_\sigma \ln Z_0(r_\uparrow, r_\downarrow) \tag{37}$$

The blocks of 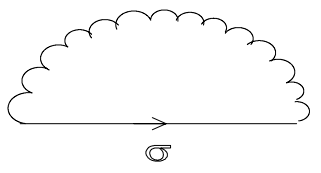  and 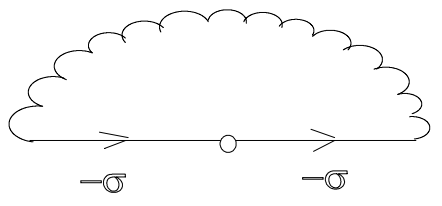  in Fig. 2 are described by the functions $\beta\delta\mu_\sigma$ and $\nu_{-\sigma} < F^{-\sigma 0} >$ :

*a)* *First approximation*

$$\delta\mu_\sigma = \frac{1}{N}\sum_{\mathbf{q}} t(\mathbf{q}) P_\sigma \sum_{m=-\infty}^{\infty} \varphi_{m\sigma} f(E_{m\sigma\mathbf{q}}) \qquad (38)$$

$$\nu_\sigma < F^{\sigma 0} > = \frac{1}{N}\sum_{\mathbf{q}} P_\sigma \sum_{m=-\infty}^{\infty} \varphi_{m\sigma} \left( f(E_{m\sigma\mathbf{q}}) - f(-m\omega_0 - \varepsilon_{0\sigma}) \right)$$

In the sums of Eq.(38) the $m_1$ and $m_2$ indices were taken equal to m, since the corresponding series terms are proportional to a high-order power of the parameter $P_\sigma\varphi_{m\sigma}$ .

*b)* *Second approximation.*

$$\delta\mu_\sigma = \frac{1}{2N}\sum_{\mathbf{q}} t(\mathbf{q}) \left\{ f(E_{1\sigma\mathbf{q}}) + f(E_{2\sigma\mathbf{q}}) + t(\mathbf{q}) < F^{\sigma 0} > \frac{f(E_{1\sigma\mathbf{q}}) - f(E_{2\sigma\mathbf{q}})}{E_{1\sigma\mathbf{q}} - E_{2\sigma\mathbf{q}}} \right\}$$

(39)

$$\nu_\sigma < F^{\sigma 0} > = \frac{1}{8N}\sum_{\mathbf{q}} \left[ \frac{t(\mathbf{q}) < F^{\sigma 0} >}{E_{1\sigma\mathbf{q}} - E_{2\sigma\mathbf{q}}} \left\{ f(E_{1\sigma\mathbf{q}}) - f(E_{2\sigma\mathbf{q}}) \right\} + f(E_{1\sigma\mathbf{q}}) + f(E_{2\sigma\mathbf{q}}) \right] -$$
$$- \frac{1}{8}\left\{ f(\xi - \varepsilon_{0\sigma}) + f(-\xi - \varepsilon_{0\sigma}) \right\}$$

Here, $E_{1\sigma\mathbf{q}}$ and $E_{2\sigma\mathbf{q}}$ are determined by the formula (35).

In Fig.2a, graphic series may be easily summed, since all terms of the sum with the exception of the last one $\nu_{-\sigma} < F^{-\sigma 0} >$ generate the Taylor power series

$$< F^{\sigma 0} >_0 - \partial_\sigma < F^{\sigma 0} >_0 \beta\delta\mu_\sigma - \partial_{-\sigma} < F^{-\sigma 0} >_0 \beta\delta\mu_{-\sigma} +$$
$$+ \frac{1}{2!}\partial_\sigma^2 < F^{\sigma 0} >_0 \left(\beta\delta\mu_\sigma\right)^2 + \frac{1}{2!}\partial_{-\sigma}^2 < F^{-\sigma 0} >_0 \left(\beta\delta\mu_{-\sigma}\right)^2 - \ldots ,$$

the sum of which is equal to $< F^{\sigma 0} >_1$ :

$$< F^{\sigma 0} >_1 = < F^{\sigma 0}(\tilde{\varepsilon}_1, \tilde{\varepsilon}_2, \tilde{\varepsilon}_3, \tilde{\varepsilon}_4, \tilde{\varepsilon}_5, \tilde{\varepsilon}_9, \tilde{\varepsilon}_{10}, \tilde{\varepsilon}_{11}, \tilde{\varepsilon}_{12}) >_0 , \qquad (40)$$

The function $< F^{\sigma 0}(\varepsilon_1,\varepsilon_2,\varepsilon_3,\varepsilon_4,\varepsilon_5,\varepsilon_9,\varepsilon_{10},\varepsilon_{11},\varepsilon_{12}) >_0$ is determined by Eq. (22) and

$$\tilde{\varepsilon}_1 = \varepsilon_1 + \delta\mu_\uparrow, \tilde{\varepsilon}_2 = \varepsilon_2 + \frac{3}{4}\delta\mu_\uparrow + \frac{1}{4}\delta\mu_\downarrow, \tilde{\varepsilon}_3 = \varepsilon_2 + \frac{1}{2}\left(\delta\mu_\uparrow + \delta\mu_\downarrow\right), \tilde{\varepsilon}_4 = \varepsilon_4 + \frac{1}{4}\delta\mu_\uparrow + \frac{3}{4}\delta\mu_\downarrow,$$

$$\tilde{\varepsilon}_5 = \varepsilon_5 + \delta\mu_\downarrow, \tilde{\varepsilon}_9 = \varepsilon_9 + \delta\mu_\uparrow + \frac{1}{4}\delta\mu_\downarrow, \tilde{\varepsilon}_{10} = \varepsilon_{10} + \frac{3}{4}\delta\mu_\uparrow + \frac{1}{2}\delta\mu_\downarrow, \tilde{\varepsilon}_{11} = \varepsilon_{11} + \frac{1}{2}\delta\mu_\uparrow + \frac{3}{4}\delta\mu_\downarrow,$$

$$\tilde{\varepsilon}_{12} = \varepsilon_{12} + \frac{1}{4}\delta\mu_\uparrow + \delta\mu_\downarrow$$

a)

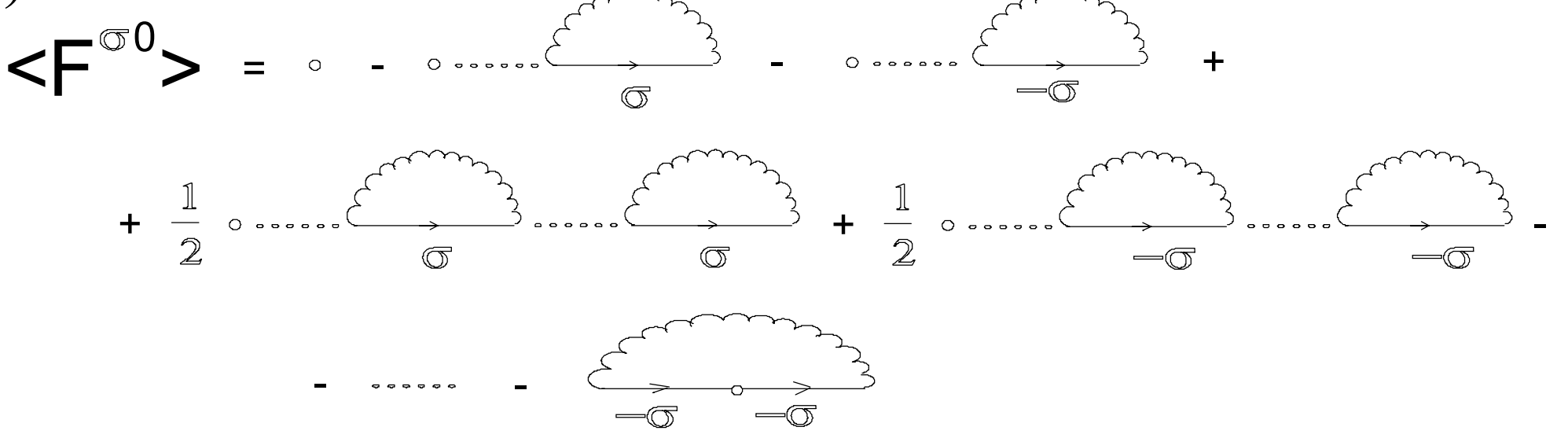

b)

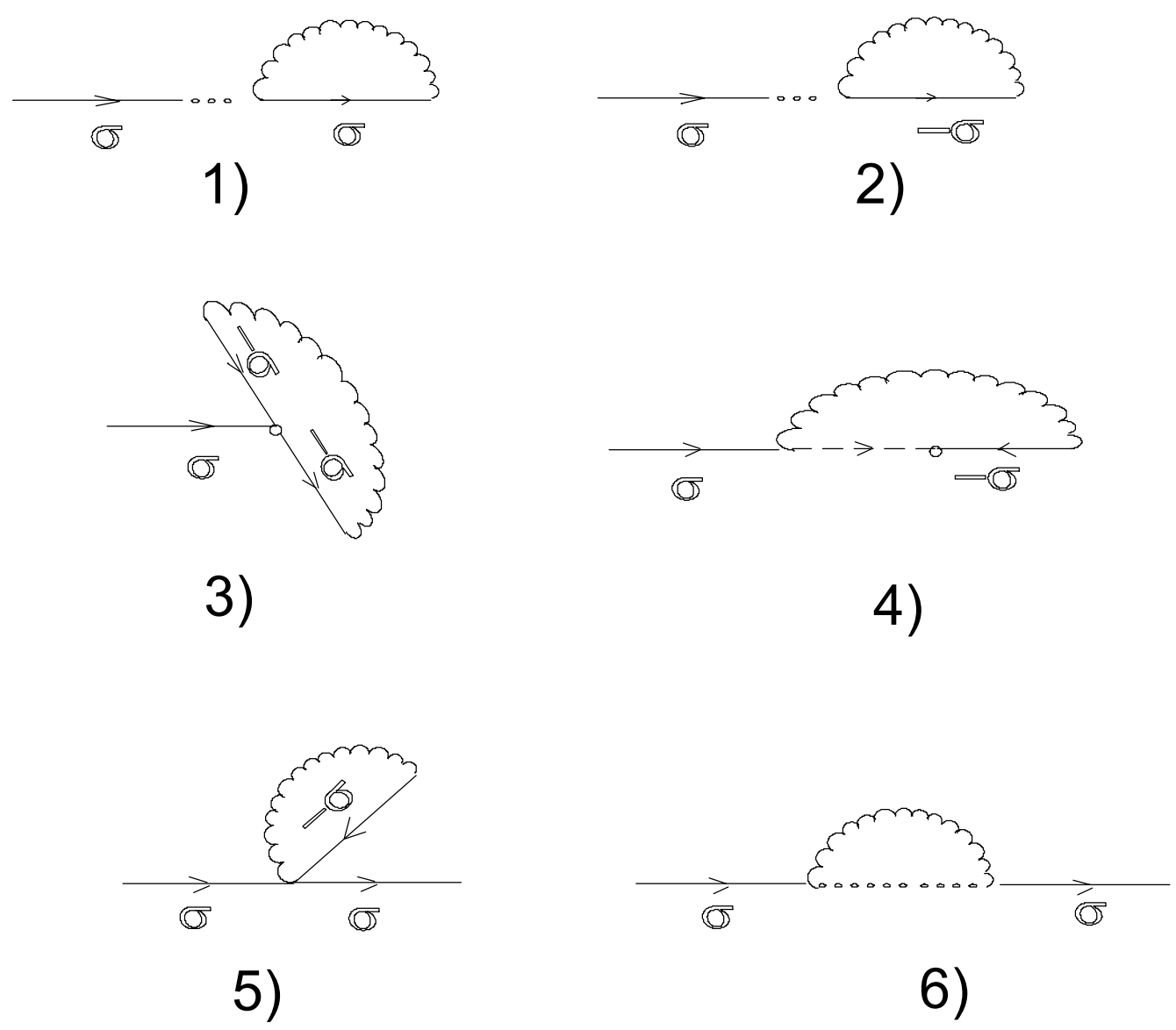

Fig.2. Graphic image of equation for combined occupancy $< F^{\sigma 0} >$ (a) and diagrams for the Green function in the first approximation of the perturbation theory (b).

Equations for the mean spin $<\sigma^z + \frac{1}{4}S_0^z>$ as well as for chemical potential μ can be written in the following form

$$\begin{aligned} \frac{1}{8}(5-n) + <\sigma^z + \frac{1}{4}S_0^z> &= <F^{+0}>_1 - \nu_{\downarrow} <F^{-0}> \\ \frac{1}{8}(5-n) - <\sigma^z + \frac{1}{4}S_0^z> &= <F^{-0}>_1 - \nu_{\uparrow} <F^{+0}> \end{aligned} \qquad (41)$$

We will consider the different solutions of the set of equations (41). Supposing that ξ = 0, T = 0 (pure double exchange) and μ > 0, we introduce the functions $E(x) = \frac{1}{6}\int_{-3}^{x} D_c(x)xdx$ и $I(x) = \int_{-3}^{x} D_c(x)dx$, where the electron density of states may be written as

$$D_c(x) = \frac{1}{N}\sum_{\mathbf{q}}\delta\left(x - \frac{t(\mathbf{q})}{2t}\right) \qquad (42)$$

From Eq. (42) it follows that for simple cubic (s.c.) lattice the variable $|x| \le 3$. The plots of *I(x)* and *E(x)* functions for this lattice are displayed in Fig. 3. Function *E(x)* characterizes an effective field contribution to electron dynamics which does not exceed 1/(2z). This contribution roughly determines the Curie temperature, $T_C$, in the units of bandwidth W.

For the case of $<F^{+0}>_1 = 1, <F^{-0}>_1 = 0$, the set of equations (41) can be written as :

$$<F^{\sigma 0}> = <F^{\sigma 0}>_1 + \frac{1}{4}\left[1 - I\left(\frac{6\mu}{W<F^{-\sigma 0}>}\right)\right], \qquad (43)$$

with solutions existing if

$$\frac{6\mu}{W<F^{-0}>} \ge 3, \qquad (44)$$

Then the expressions for both mean spin of saturated ferromagnet (FM) and chemical potential have the form $<\sigma^z + \frac{1}{4}S_0^z> = \frac{1}{8}(3+n)$ and $\frac{\tilde{\mu}}{W} = \frac{1}{6}I^{-1}(n)$, respectively, where $I^{-1}(x)$ is the inverse function of $I(x)$. In Fig. 3 it is seen that $I^{-1}(x) > 0$ for x > 0.5. Therefore, the self-consistent FM solution is valid for electron concentrations *n > 1/2*. In perovskite manganites this fact was fixed experimentally in work [17]. Using Eq. (44) we obtain a more precise estimation of electron concentration, namely, $n_{FM} \approx 0.588$ above which the FM state exists.

The set of Eq. (41) has solutions $<\sigma^z + \frac{1}{4}S_0^z> =0$, $\tilde{\mu} = \tilde{\mu}_{PM2} = \frac{W}{48}(5-n)I^{-1}\left(\frac{n+1}{2}\right)$ and $\tilde{\mu} = \tilde{\mu}_{PM1} = \frac{W}{48}(5-n)I^{-1}\left(\frac{n}{2}\right)$, corresponding to two paramagnetic phases PM-2 and PM-1. In these phases $<F^{\sigma 0}>_1 = 1/2, \tilde{\mu} + \delta\mu_\sigma/4 > 0$ and $<F^{\sigma 0}>_1 = 5/8, \tilde{\mu} + \delta\mu_\sigma/4 < 0$, respectively. The existence of two paramagnetic phases was also predicted in work [8] for HTSC systems.

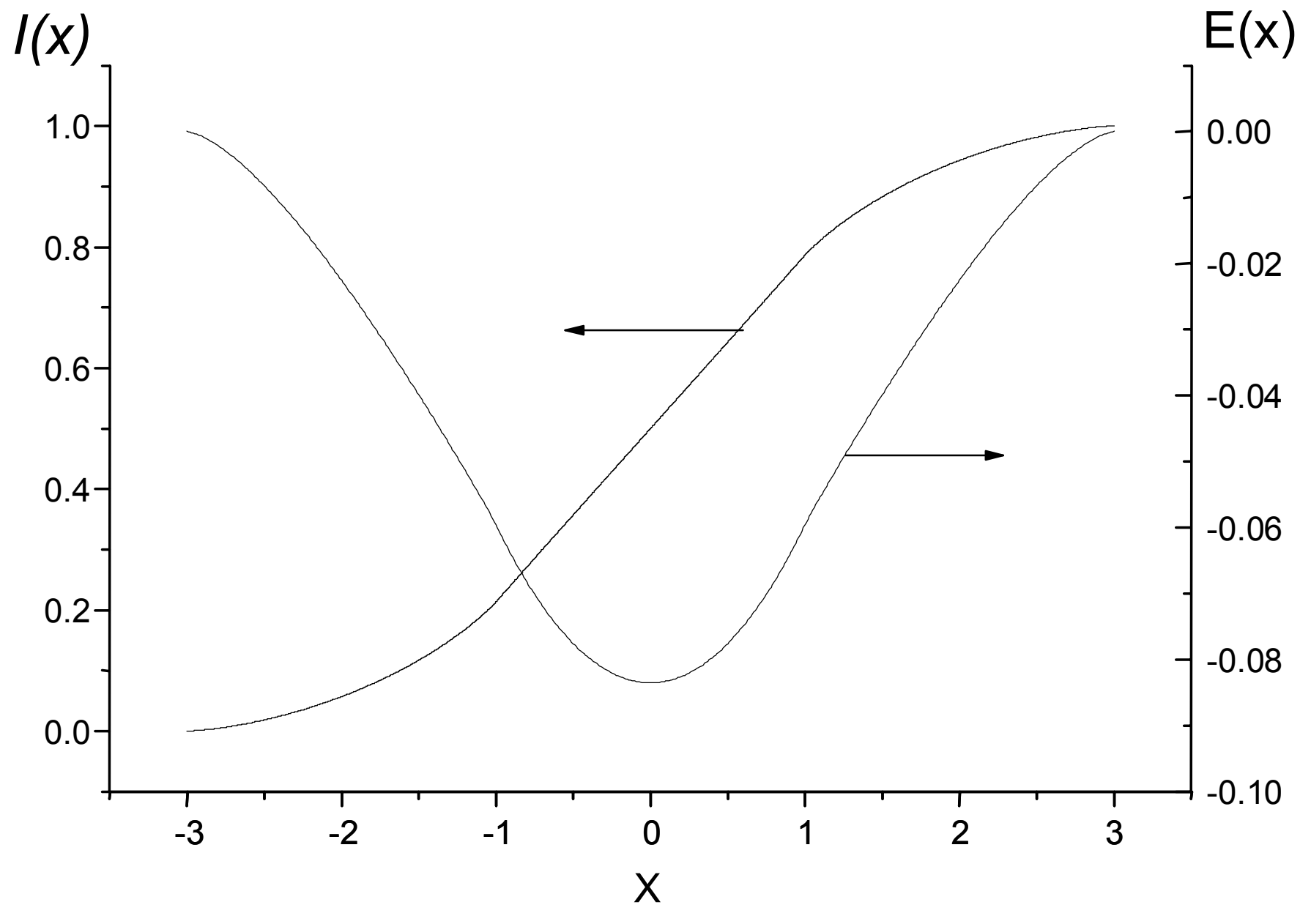

Fig.3. *I(x)* and *E(x)* functions.

To find the temperature $T_C$ the set of equations in the linear approximation is expanded in terms of small parameter $<\sigma^z+\frac{1}{4}S_0^z>$ at T ~ $T_C$. Neglecting superexchange and supposing $\xi$ = 0 (pure DE model) we obtain the following equations for $\mu$ and $T_C$:

$$1-n=\frac{1}{1+\frac{5}{4}e^{\beta_C(\tilde{\mu}+\frac{1}{4}\delta\mu)}}-\frac{2}{N}\sum_{\mathbf{q}}f(E_{0\mathbf{q}})+2f(-\tilde{\mu})$$

$$4T_C^2=\frac{1}{N}\sum_{\mathbf{q}}t(\mathbf{q})f(E_{0\mathbf{q}})\left[1-f(E_{0\mathbf{q}})\right]\left\{5t(\mathbf{q})\frac{e^{\beta_C(\tilde{\mu}+\frac{1}{4}\delta\mu)}+\frac{1}{2}}{5e^{\beta_C(\tilde{\mu}+\frac{1}{4}\delta\mu)}+4}-T_C\right\}, \qquad (45)$$

where $\beta_C=1/T_C$, $\delta\mu=\frac{1}{N}\sum_{\mathbf{q}}t(\mathbf{q})f(E_{0\mathbf{q}})$, $E_{0\mathbf{q}}=-\tilde{\mu}+\frac{1}{8}(5-n)t(\mathbf{q})$.

The results of $T_C(n)$ calculation in W units are shown in Fig. 4 (curve 1). The dependence $T_C(n)$ obtained assuming that $\tilde{\mu}=\tilde{\mu}_{PM2}$ (see the above formula for chemical potential defined at T = 0) and $\frac{df(E_{0\mathbf{q}})}{dE_{0\mathbf{q}}}=-\delta(E_{0\mathbf{q}})$ has the form :

$$T_C/W\approx\frac{768(\tilde{\mu}/W)^2D_c\left(\frac{48\tilde{\mu}/W}{5-n}\right)}{\left[(5-n)\right]^3+384(5-n)D_c\left(\frac{48\tilde{\mu}/W}{5-n}\right)\tilde{\mu}/W} \qquad (46)$$

and is also presented in Fig. 4 (curve 2). Formula (46) somewhat differs from the result obtained in Ref. 9, when Hund's coupling was considered in the mean field approximation. The curve (3) corresponds to approximate solution of the set of equations (45) with chemical potential at T = 0 and step Fermi function (Eq. (46)). Therefore, one can note that rigorous account of quantum spin fluctuation with Hund's coupling decreases the Curie temperature essentially (up to 30 %).

The above equations correspond to pure DE model. There are no difficulties in derivation of similar formulae taking into account the electron-phonon coupling. Unfortunately, either of the two above considered approximations gives different results for magnetic and transport properties.

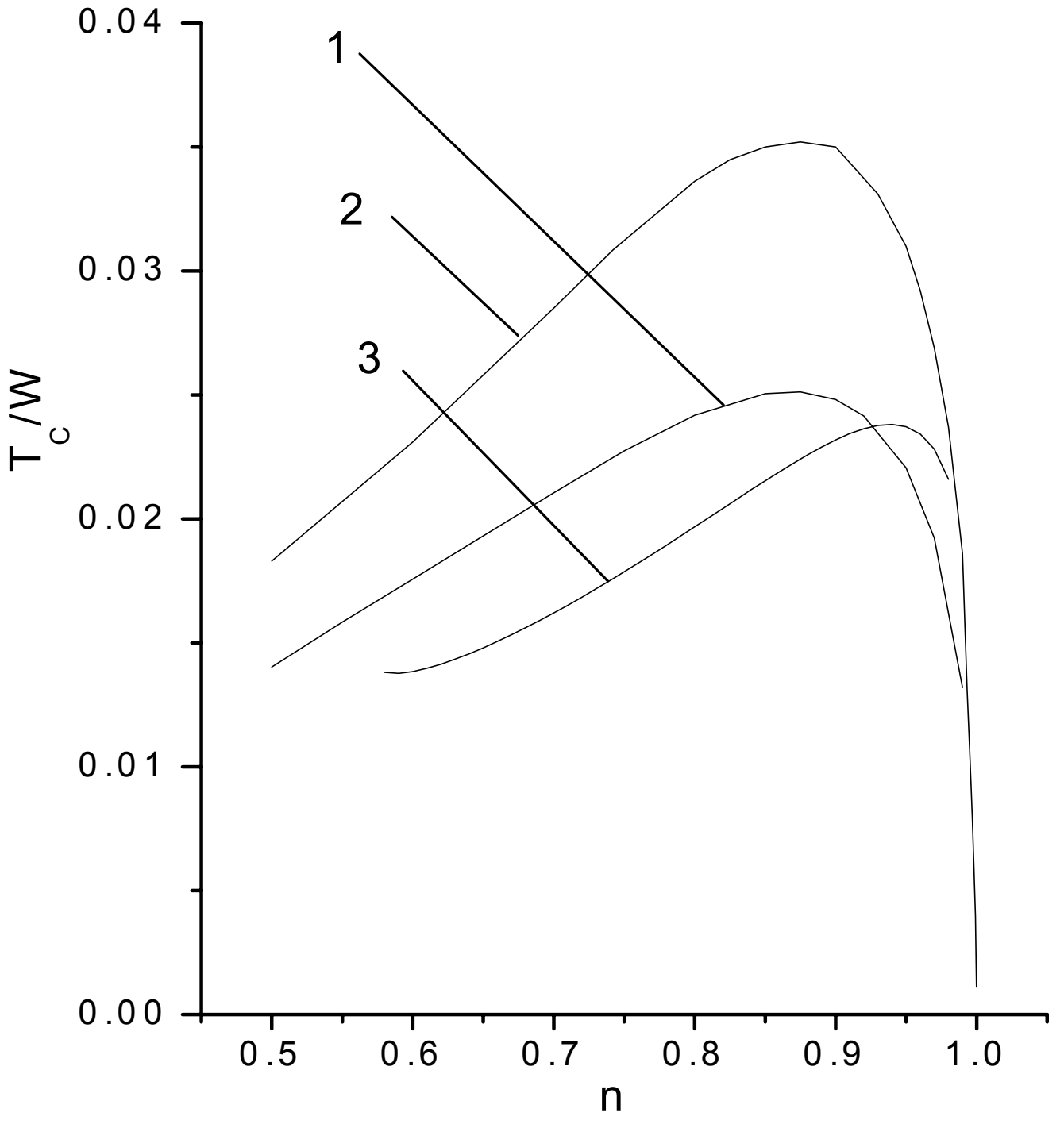

Fig.4. The Curie temperature, $T_C$, versus electron concentration, n, with account of Hund's rule coupling in the framework of given theory (curve 1), in the mean field approximation [8] (curve 2). Curve (3) corresponds to approximate solution of the set of equations (45).

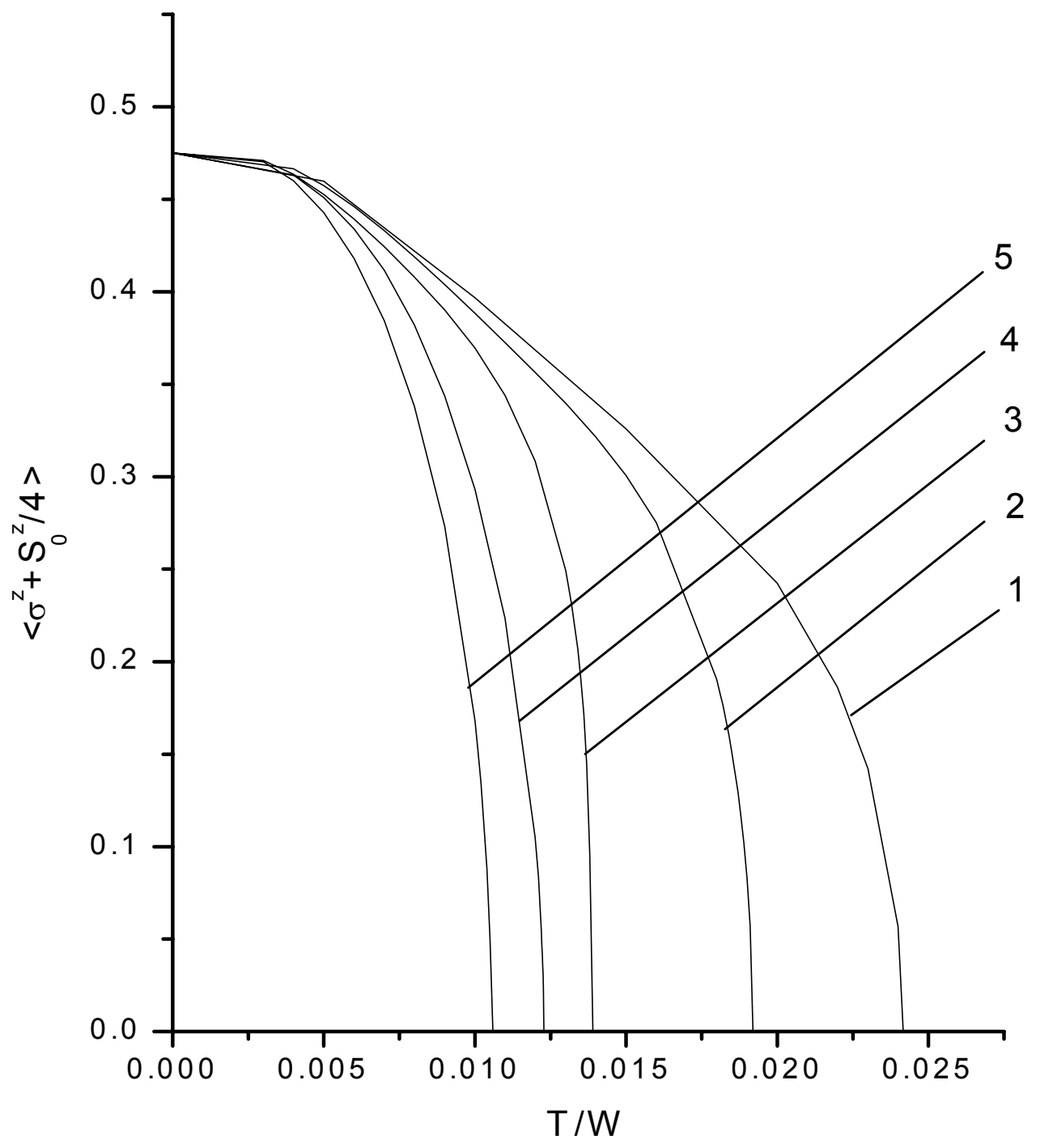

Fig.5. Temperature dependences of magnetization $<\sigma^z + \frac{1}{4} S_0^z>$ for n = 0.8 with $\xi$ / W = 0, 0.003, 0.01, 0.03 and 0.04 (curves 1-5, respectively) .

In the first approximation (Eq. 38) the strong influence of polaron binding energy $\xi$ on the ferromagnetic phase is observed. The temperature dependences of mean spin obtained solving the set of equations (36) at electron concentration n = 0.8 for various values of $\xi$ are illustrated in Fig. 5. Hereinafter the phonon frequency value $\omega_0 / W = 0.025$ and the lattice constant a = 5Å were used. It is seen that the Curie temperature is reduced more than by half at $\xi$/W = 0.04. The chemical potential also decreases abruptly. It is indicative of narrowing effective band. The size of polaron calculated using Holstein expression aW / $\xi \approx$ 25a is not small. Therefore, the electron-phonon interaction suppresses DE interaction as predicted in [6,10].

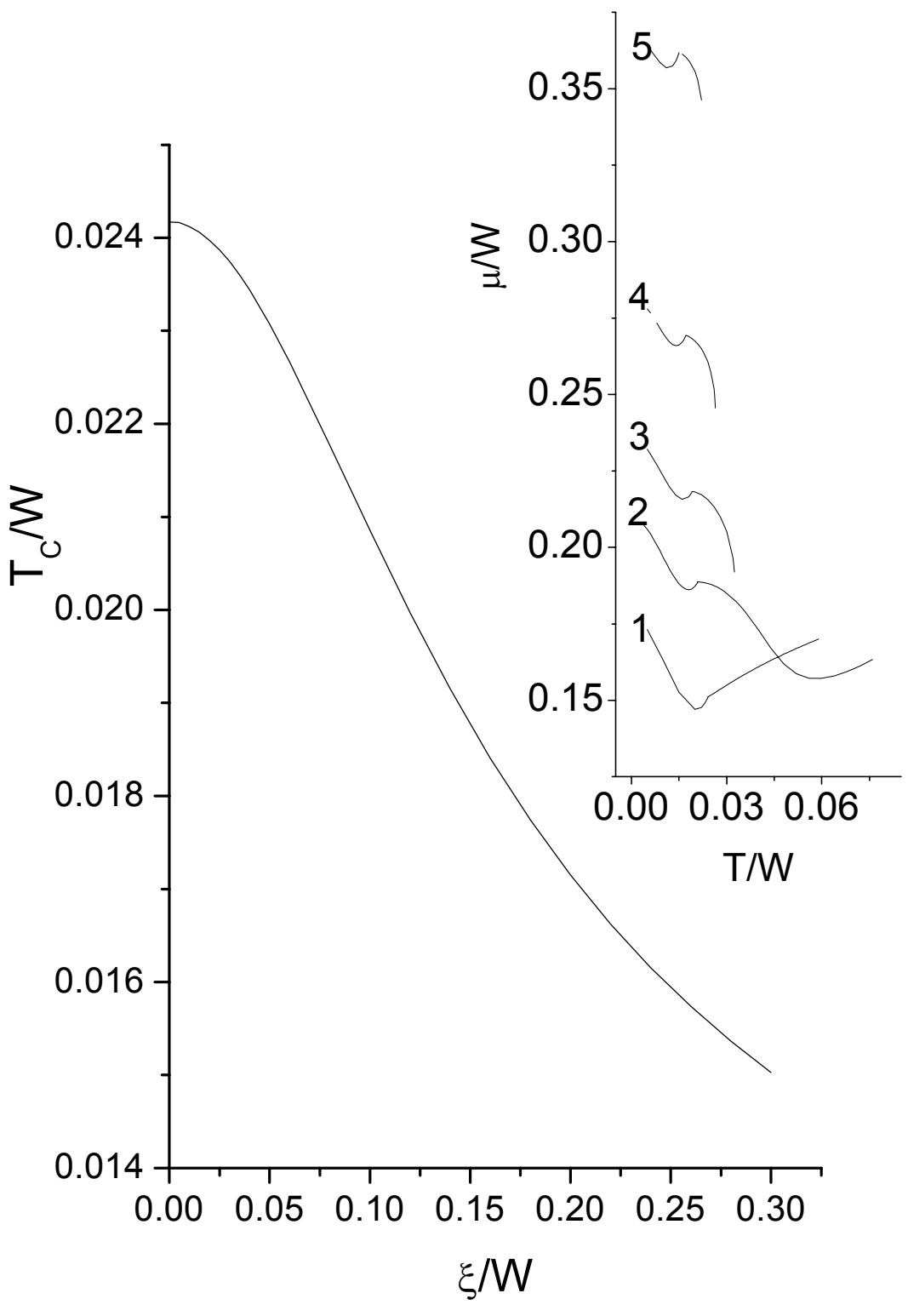

Fig.6. The Curie temperature, $T_C$, versus polaron energy $\xi$ at n = 0.8. Inset : the temperature dependences of the chemical potential $\mu$ at $\xi$ / W = 0, 0.1, 0.14, 0.2 and 0.3 ( curves 1-5, respectively).

In Fig. 6 the calculated in the second approximation (35) dependence $T_C(\xi)$ is shown. One can see that $T_C$ is decreased by one third at $\xi/W \sim 0.25$ that corresponds to small polaron theory. However, in this case the chemical potential is increased (see inset of Fig. 6). The increase of $\mu$ near the phase transition with increasing polarons binding energy reflects the processes of polaron destruction owing to scattering on spin fluctuations. Note that the function $T_C(g)$ obtained in Ref.6 decreases faster with increasing g parameter at n = 0.5. Therefore, the contribution of electron correlations is expected to be overestimated in CPA method.

Fig. 7. shows the influence of both superexchange and applied magnetic field, h, on magnetic ordering temperature with account of electron-phonon coupling (35). The value J(0)/W = 0.00215 obtained in [18] for $LaMnO_3$ corresponds to J(0) = 25 K. The value h/W = 0.0001 will correspond to applied magnetic field of 8.64 kOe, since the bandwidth W = 1 eV. In Fig. 7 is seen that superexchange increases $T_C$ while applied magnetic field smears the phase transition as it has to be in the 2-nd kind phase transitions theory.

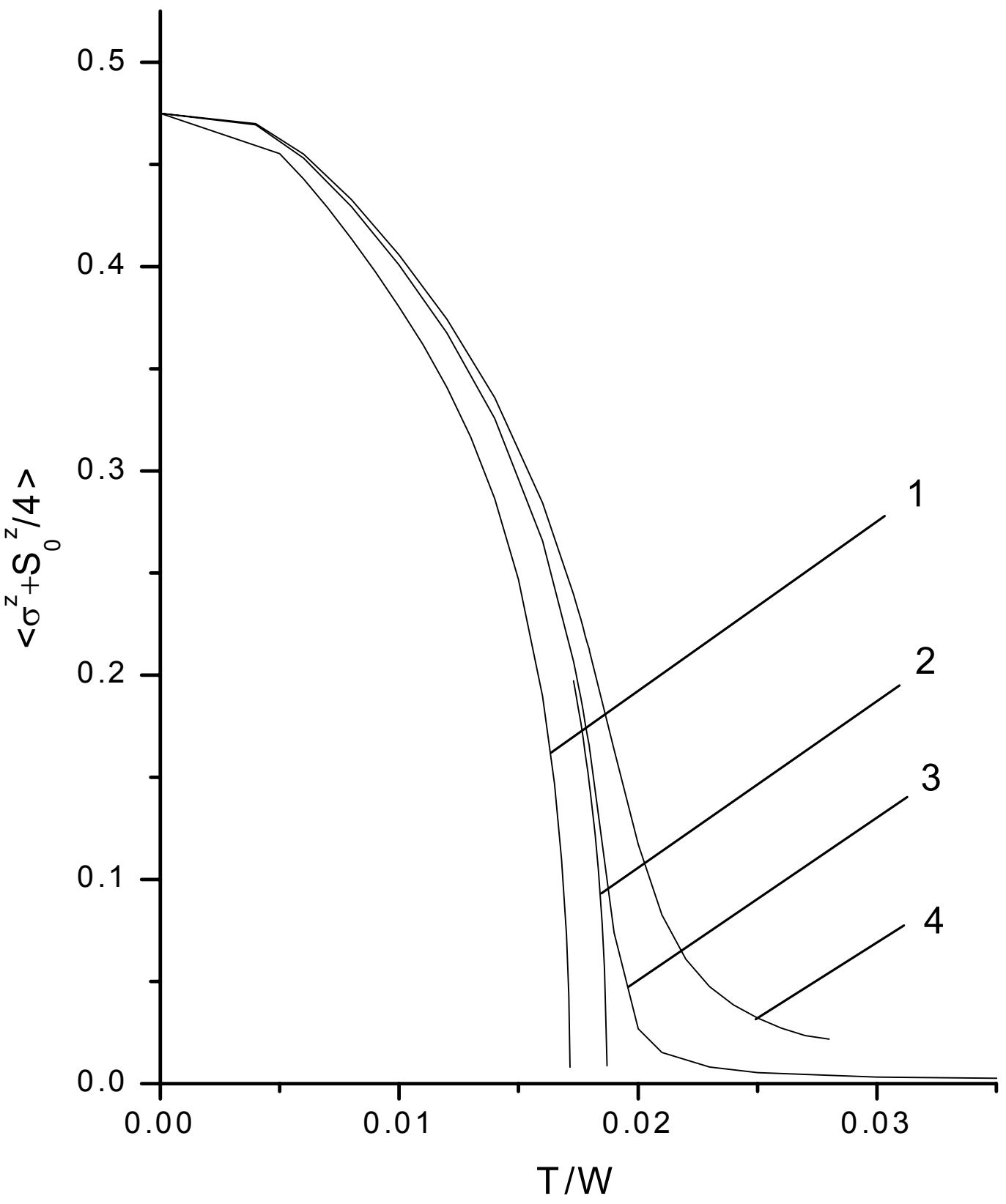

Fig. 7. Influence of both applied magnetic field, h, and superexchange, J(0), on the temperature dependences of magnetization at n = 0.8 and ξ/W = 0.2 : 1) J(0) = 0, h = 0 and J(0)/W = 0.00215 for 2) h = 0; 3) h/W = 0.0001 and 4) h/W = 0.0006.

## 6. Spectral and transport properties of electron-hole excitations.

In Fig. 2b, all diagrams in the first order relatively to inverse effective radius of interaction for the Green's function $\mathcal{G}_\sigma(i\omega_n,\mathbf{k})$ are shown. Diagrams 1, 2, 3 and similar in a higher order of expansion forming a series $<F^{\sigma 0}>$ are depicted in Fig. 2a. Consequently, the appropriate correction to $\mathcal{G}_\sigma(i\omega_n,\mathbf{k})$ is equal to $<F^{\sigma 0}> G_{0\sigma}(i\omega_n)$.

Let us write the analytic expressions for diagrams 4, 5 and 6 in the first approximation (31) for electron-phonon interaction:

$$\Lambda_4^\sigma(i\omega_n)=\frac{\sigma}{2N}\sum_{\mathbf{q}m}\frac{t(\mathbf{q})P_{-\sigma}\varphi_{m-\sigma}P_0\varphi_{00}}{i\omega_n-E_{m-\sigma\mathbf{q}}+\varepsilon_{-\sigma\sigma}}\left\{f(E_{m-\sigma\mathbf{q}})+b(\varepsilon_{-\sigma\sigma})\right\}<\sigma^z+\frac{1}{4}S_0^z>_0$$

$$\Delta_5^\sigma=-\frac{1}{4}\beta\delta\mu_{-\sigma}<F^{\sigma 0}> \qquad (47)$$

$$\Delta_6^\sigma(i\omega_n)=\frac{1}{N}\sum_{\mathbf{q}m}\frac{\beta t^2(\mathbf{q})<F^{\sigma 0}>P_\sigma\varphi_{m\sigma}}{i\omega_n-E_{m\sigma\mathbf{q}}}<F_p^{\sigma 0}F_l^{\sigma 0}>_0$$

Here, *b(x)=1/(exp(βx)-1)* is the Bose distribution function, $P_0\varphi_{00}=e^{-\lambda^2(2B+1)}I_0\left(2\lambda^2\sqrt{B(B+1)}\right)$, $<\sigma^z+\frac{1}{4}S_0^z>_0\approx\frac{5}{4}\beta\tilde{H}\frac{2e^{\beta\tilde{\mu}}+1}{5e^{\beta\tilde{\mu}}+4}$. Since $\beta\tilde{H}\ll 1$, the nonzero contribution to $\Lambda_4^\sigma(i\omega_n)$ is proportional to $b(\varepsilon_{-\sigma\sigma})<\sigma^z+\frac{1}{4}S_0^z>_0\approx\frac{1}{2\sigma}$ at $\beta\tilde{\mu}\gg 1$. Also, for linked diagrams which are proportional to $\delta_{pl}$ we have $<F_p^{\sigma 0}F_l^{\sigma 0}>_0=m(\tilde{\mu})\delta_{pl}$ where $m(\tilde{\mu})=\frac{5}{8}\frac{(5e^{\beta\tilde{\mu}}+2)(e^{\beta\tilde{\mu}}+1)}{(5e^{\beta\tilde{\mu}}+4)^2}$.

The formulae for diagrams 4, 5 and 6 in a second approximation (34)-(35) can be presented in the form :

$$\Lambda_4^{-\sigma}(i\omega_n) = \frac{-\sigma}{4N}\sum_{\mathbf{q}} \frac{t(\mathbf{q}) < \sigma^z + \frac{1}{4}S_0^z >_0}{E_{1\sigma\mathbf{q}} - E_{2\sigma\mathbf{q}}} \begin{Bmatrix} \frac{E_{1\sigma\mathbf{q}} + \varepsilon_{0\sigma}}{i\omega_n - E_{1\sigma\mathbf{q}} + \sigma\mu_1}\left[f(E_{1\sigma\mathbf{q}}) + b(\sigma\mu_1)\right] + \\ \frac{E_{1\sigma\mathbf{q}} + \varepsilon_{0\sigma}}{i\omega_n - E_{1\sigma\mathbf{q}} + \sigma\mu_2}\left[f(E_{1\sigma\mathbf{q}}) + b(\sigma\mu_2)\right] - \\ \frac{E_{2\sigma\mathbf{q}} + \varepsilon_{0\sigma}}{i\omega_n - E_{2\sigma\mathbf{q}} + \sigma\mu_1}\left[f(E_{2\sigma\mathbf{q}}) + b(\sigma\mu_1)\right] - \\ \frac{E_{2\sigma\mathbf{q}} + \varepsilon_{0\sigma}}{i\omega_n - E_{2\sigma\mathbf{q}} + \sigma\mu_2}\left[f(E_{2\sigma\mathbf{q}}) + b(\sigma\mu_2)\right] \end{Bmatrix}$$

$$\Delta_6^{\sigma}(i\omega_n) = \frac{1}{2N}\sum_{\mathbf{q}} \frac{t^2(\mathbf{q}) < F^{\sigma 0} >}{E_{1\sigma\mathbf{q}} - E_{2\sigma\mathbf{q}}} \left\{ \frac{E_{1\sigma\mathbf{q}} + \varepsilon_{0\sigma}}{i\omega_n - E_{1\sigma\mathbf{q}}} - \frac{E_{2\sigma\mathbf{q}} + \varepsilon_{0\sigma}}{i\omega_n - E_{2\sigma\mathbf{q}}} \right\} < F_p^{\sigma 0} F_l^{\sigma 0} >_0 \qquad (48)$$

The expression for $\Delta_5^{\sigma}$ is identical to one presented in Eq. (47) having in mind that $\delta\mu_{\sigma}$ and $E_{i\sigma\mathbf{q}}$ are determined by Eqs. (35) and (39), respectively. We will introduce $\mu_1 = \varepsilon_{\uparrow\downarrow} + \xi$ , $\mu_2 = \varepsilon_{\uparrow\downarrow} - \xi$ . Since the parameters $\mu_i$ are finite at $\tilde{H} \to 0$, the product $< \sigma^z + \frac{1}{4}S_0^z >_0 b(\sigma\mu_i) \to 0$. Then the contribution $\Lambda_4^{-\sigma}(i\omega_n)$ can be neglected.

To find the final expressions for diagrams, the value of $\Lambda_4^{\sigma}(i\omega_n)$ should be multiplied by $G_{0\sigma}(i\omega_n)$ and $\Delta_5^{\sigma}$ и $\Delta_6^{\sigma}$ by $G_{0\sigma}(i\omega_n)^2$. Then self-energy $\sum_{\sigma}(i\omega_n)$ is written as

$$\sum\nolimits_{\sigma}(i\omega_n) = < F^{\sigma 0} > G_{0\sigma}(i\omega_n) + \Lambda_4^{\sigma}(i\omega_n) G_{0\sigma}(i\omega_n) + (\Delta_5^{\sigma} + \Delta_6^{\sigma}(i\omega_n)) G_{0\sigma}(i\omega_n)^2 \qquad (49)$$

Using Larkin equation [19] the total Green's function $\mathcal{G}_{\sigma}(i\omega_n, \mathbf{k})$ can be easily found:

$$\mathcal{G}_{\sigma}(i\omega_n, \mathbf{k}) = \frac{\sum_{\sigma}(i\omega_n)}{1 - \beta t(\mathbf{k}) \sum_{\sigma}(i\omega_n)} \qquad (50)$$

In DE model ($\xi = 0$) this formula is essentially simplified:

$$\mathcal{G}_{\sigma}(i\omega_n, \mathbf{k}) = \frac{1}{\beta} \frac{< F^{\sigma 0} > + \Lambda_4^{\sigma}(i\omega_n)}{i\omega_n - \omega_{\sigma\mathbf{k}}}, \qquad (51)$$

where $\omega_{\sigma\mathbf{k}} = -\varepsilon_{0\sigma} - \frac{1}{4}\delta\mu_{-\sigma} + \left(< F^{\sigma 0} > + \Lambda_4^{\sigma}(i\omega_n)\right)t(\mathbf{k}) + \frac{\Delta_6^{\sigma}(i\omega_n)}{\beta < F^{\sigma 0} >}$.

To deduce Eq. (51) we have used linear expansion of small parameters. Realizing in Eqs. (50) and (51) an analytical continuation $i\omega_n \to \omega + i\delta$ we obtain retarded Green's function the poles of which determine the $\Omega(\mathbf{k})$ spectrum of excitations. Imaginary part $\mathcal{G}_{\sigma}(i\omega_n, \mathbf{k})$ determines the spectral density being proportional to $\delta(\omega - \Omega(\mathbf{k}))$ for coherent excitations.

The spectral density of incoherent spectrum describing the relaxation processes is of principal interest. After analytical continuation $i\omega_n \to \omega + i\delta$ using Eqs. (47) and (48) one can evaluate the nonzero imaginary parts of $\Lambda_4^{\sigma}(\Omega(\mathbf{k}))$ и $\Delta_6^{\sigma}(\Omega(\mathbf{k}))$. The corresponding formulae are given in Appendix (Eqs. (A.5) and (A.6)). $\sum_{\sigma}{}'(\omega)$ and $\sum_{\sigma}{}''(\omega)$ correspond to real and imaginary parts of self-energy $\sum_{\sigma}(\omega)$, respectively. Let us write the imaginary part $\mathrm{Im}(\mathcal{G}_{\sigma}(i\omega_n, \mathbf{k}))$ of total Green's function as

$$\mathrm{Im}(\mathcal{G}_{\sigma}(\omega + i\delta, \mathbf{k})) = \frac{\sum_{\sigma}{}''(\omega)}{\left(1 - \beta t(\mathbf{k})\sum_{\sigma}{}'(\omega)\right)^2 + \left(\beta t(\mathbf{k})\sum_{\sigma}{}''(\omega)\right)^2} \qquad (52)$$

In the case of incoherent excitations the spectral density $R_{\sigma}(\omega, \mathbf{k}) = -2\beta\, \mathrm{Im}(\mathcal{G}_{\sigma}(\omega + i\delta, \mathbf{k}))$ will be different from zero over a certain frequency interval in which the electron density of state $D_C(x)$ has also nonzero value.

As noted above, the pole singularity of Eq. (50) allows to find the dispersion law for excitation spectrum :

$$t(\mathbf{k}) = t_{0\mathbf{k}} = \mathrm{Re}\left(\frac{1}{\beta \sum_{\sigma}(\Omega(\mathbf{k}))}\right) = \frac{1}{\beta}\frac{\sum_{\sigma}{}'(\Omega(\mathbf{k}))}{\left[\sum_{\sigma}{}'(\Omega(\mathbf{k}))\right]^2 + \left[\sum_{\sigma}{}''(\Omega(\mathbf{k}))\right]^2} \qquad (53)$$

It is easy to check that derivative $\frac{dR_\sigma(\omega,\mathbf{k})}{dt(\vec{k})}$ is zero at $t(\mathbf{k}) = t_{0\mathbf{k}}$. Thus, the extremal points of spectral density as a function of hopping integral $t(\mathbf{k})$ determine $t_{0\mathbf{k}}$ corresponding to resonance frequency $\Omega(\mathbf{k})$. On the other hand, the positions of sharp peaks at the frequency curve determine excitations closely related to the coherent ones when the imaginary part of self energy is small.

In Fig. 8a,b the frequency dependences of imaginary parts of $\Lambda_4^\sigma(\omega)$ and $\Delta_6^\sigma(\omega)$ as well as spectral density of incoherent spectrum for up- and down-spin bands are shown for ξ = 0, T = 0 and n = 0.8. The peaks of $R_\sigma(\omega,\mathbf{k})$ appear near frequencies at which the imaginary parts of $\Lambda_4^\sigma(\omega)$ and $\Delta_6^\sigma(\omega)$ diagrams tend to zero. The negative values of spectral density in a narrow frequency interval indicate that because of the smallness of imaginary part the diagrams of higher order for the effective radius of interaction are to be accounted for. It should be emphasized that in this theory $R_\sigma(-\tilde{\mu},\mathbf{k}) = 0$.

In CPA method [20] the spectral density is nonzero at the Fermi level that is a serious defect since in this case the Luttinger theorem[21] for Fermi liquid is violated. In our case such a problem does not arise because the spectral density of incoherent excitations tends to zero near the Fermi level at $\omega \sim -\tilde{\mu}$.

Fig. 9 displays the frequency dependence of spectral density $R_\sigma(\omega,\vec{k})$ in the approximation (31) for effective line at n = 0.8, ξ/W = 0.03 and t(**q**)/W = 1/3. The curves 1 and 2 correspond to up- and down-spin bands at low temperature (T/W = 0.005) and curve 3 to PM-2 phase. It is seen that very sharp peaks characteristic of coherent polaron excitation are only in the PM-2 phase (see inset in Fig.9). Therefore polarons are involved in conductivity above the temperature $T_C$ and increase it. Unfortunately, the value ρ(T) near $T_C$ (in PM phase) does not exceed a few mOhm*cm that is inconsistent with the experiment. Seemingly, near the phase transition the vertex corrections in Green's function begin to play an important role. This problem is very complicated and demands special consideration.

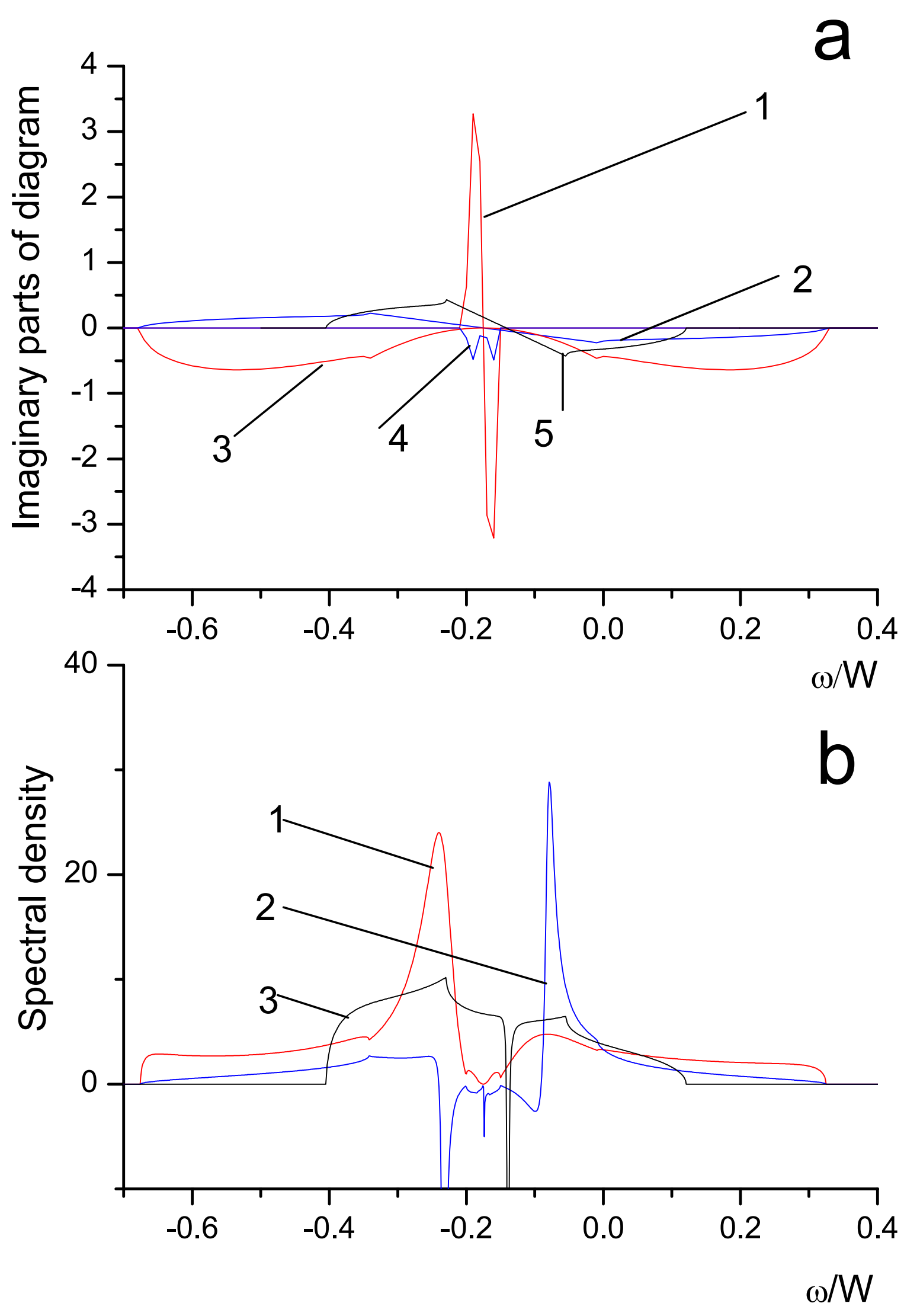

Fig. 8. At n = 0.8, $\xi = 0$ and T = 0 : a) the imaginary parts of diagrams $\Lambda_4^\sigma(\omega)$ (curves 1 and 2) and $\Delta_6^\sigma(\omega)$ (curves 3 and 4) for up- and down-spin bands, respectively. The curve 5 corresponds to $\Lambda_4^\sigma(\omega)$ in PM phase; b) spectral density as a function of frequency for up- and down-spin bands and PM phase (curves 1-3, respectively) at $t(\mathbf{q})/W = -0.1$.

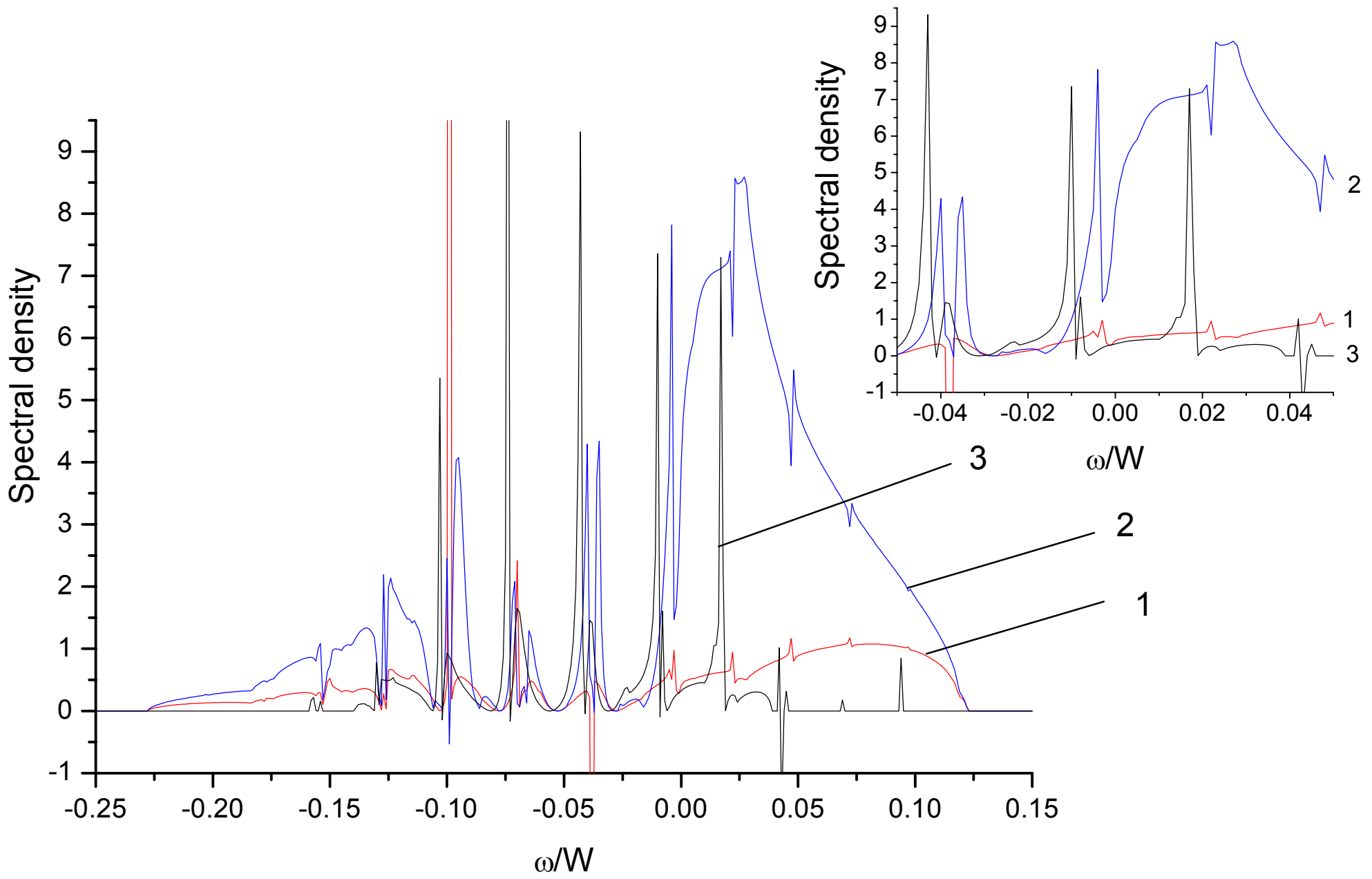

Fig. 9. Spectral density versus frequency obtained using the first approximation (Eq.(31)) at $\xi = 0.03$, $t(\mathbf{q})/W = 1/3$, n = 0.8 and T/W = 0.005 for up- and down-spin bands (curves 1 and 2, respectively) and for PM phase at T/W = 0.01228 (curve 3). Inset : the same plot near the zero frequency on an enlarged scale.

The second approximation (Eqs. (35), (39) and (48)) is applicable to describe magnetoresistive effect. One will suppose that in the excitations spectrum a gap of the order of small polaron binding energy $\xi$ is formed at $T \sim T_C$. Using Eqs. (35), (39) and (48) we obtain the ~ 30% reduction of the $T_C$ at $\xi/W = 0.2$. The chemical potential lies in the middle of a forbidden band. In Fig.10 the spectral densities as a function of frequency for up- and down-spin bands at T/W = 0.015 and for PM phase at T/W = 0.0172 are shown at $\xi/W = 0.2$, n = 0.8 and $t(\mathbf{q})/W = 0.1$. This correlates with results obtained in [6, 20]. In contrast to CPA the spectral density in Fig.10 is asymmetrical relatively frequency $\omega \sim -\tilde{\mu}$.

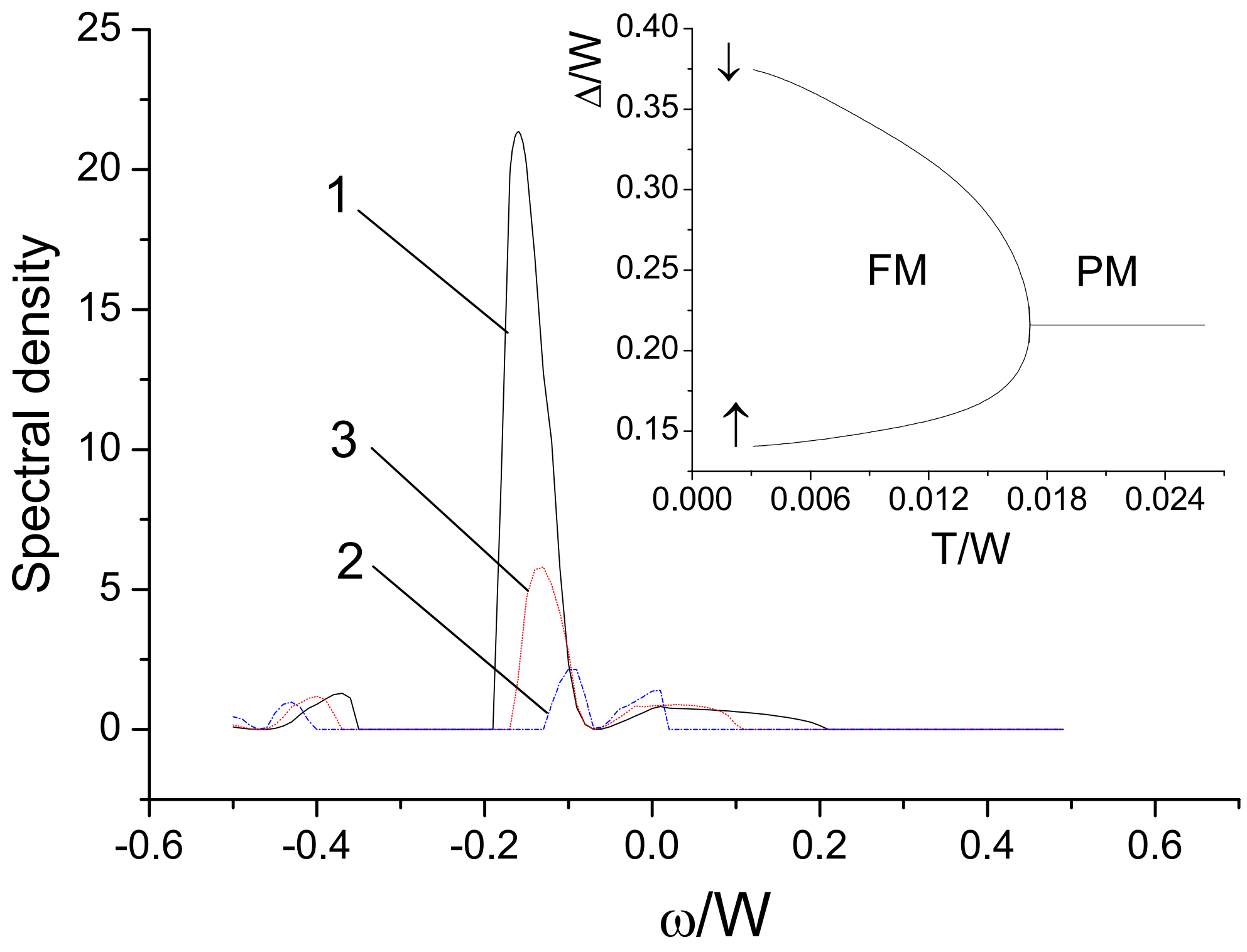

Fig. 10. Spectral density versus frequency using the second approximation (Eq.(35)) for effective line of electron-phonon interaction for up- and down-spin bands (curves 1 and 2, respectively, at T/W=0.015) and for the PM-2 phase (T/W=0.0172) at ξ=0.2, $t(\mathbf{q})/W = 0.1$, n = 0.8. Inset illustrates the temperature dependences of quasigap Δ for up-, down-spin bands and PM-2 phase.

The difference is that in the CPA method a pseudogap occurs at g/W ≈ 0.15 and the nonzero spectral density at the Fermi level increases with increasing temperature. Inset to Fig. 10 illustrates the temperature dependence of the gap for up- and down-spin bands. In this approximation the gap in PM-2 phase near the phase transition does not depend practically on temperature.

The expression for conductivity $\sigma(T)$ of s.c. lattice obtained on the basis of the above results using the Kubo formula in the bubble approximation [5] has the form :

$$\sigma(T) = \frac{e^2}{3a\pi\hbar N}\sum_{\mathbf{q}\sigma}\Phi_\sigma(t(\mathbf{q}))t(\mathbf{q}), \tag{54}$$

where the $\Phi_\sigma(t(\mathbf{q}))$ function is related to the spectral density by the following differential equation:

$$\frac{d\Phi_\sigma(t(\mathbf{q}))}{d(t(\mathbf{q}))} = \left[R_\sigma(\omega,\mathbf{q})\right]^2 \qquad (55)$$

Since the imaginary part of the Green's function (Eq. (52)) is a simple function of parameter $t(\mathbf{q})$ the integration of squared spectral density in Eq. (55) presents no problem. Here we are not presenting the sufficiently complicated expression for $\Phi_\sigma(t(\mathbf{q}))$.

In Fig. 11 the resistivity versus band filling ($\rho$(n)) in the FM and PM-2 phases (curves 1 and 2, respectively) are shown at T = 0 and $\xi$ = 0. Curve 3 in Fig. 11 shows the dependence $\rho$(n) in the PM phase calculated by the CPA method [5]. It is seen from Fig. 11 that the CPA method gives the minimal $\rho$(n) value of the order of 1 mOhm*cm. In our case the resistivity $\rho$(n) in the PM-2 and FM phases is well smaller near their lower boundaries at $n \geq n_0$ ($n_0$ is equal to 0.116 and 0.588 for the PM-2 and FM phases, respectively). According to the experimental data [22] in $La_{1-x}Sr_xMnO_3$ the residual resistance at $x \sim 0.3$ is essentially smaller than 1 mOhm*cm. At $n \to 1$ the resistivity $\rho$(n) in the FM phase abruptly increases as in the CPA method. However, for dielectric state (n = 1) the higher order diagrams must be taken into account.

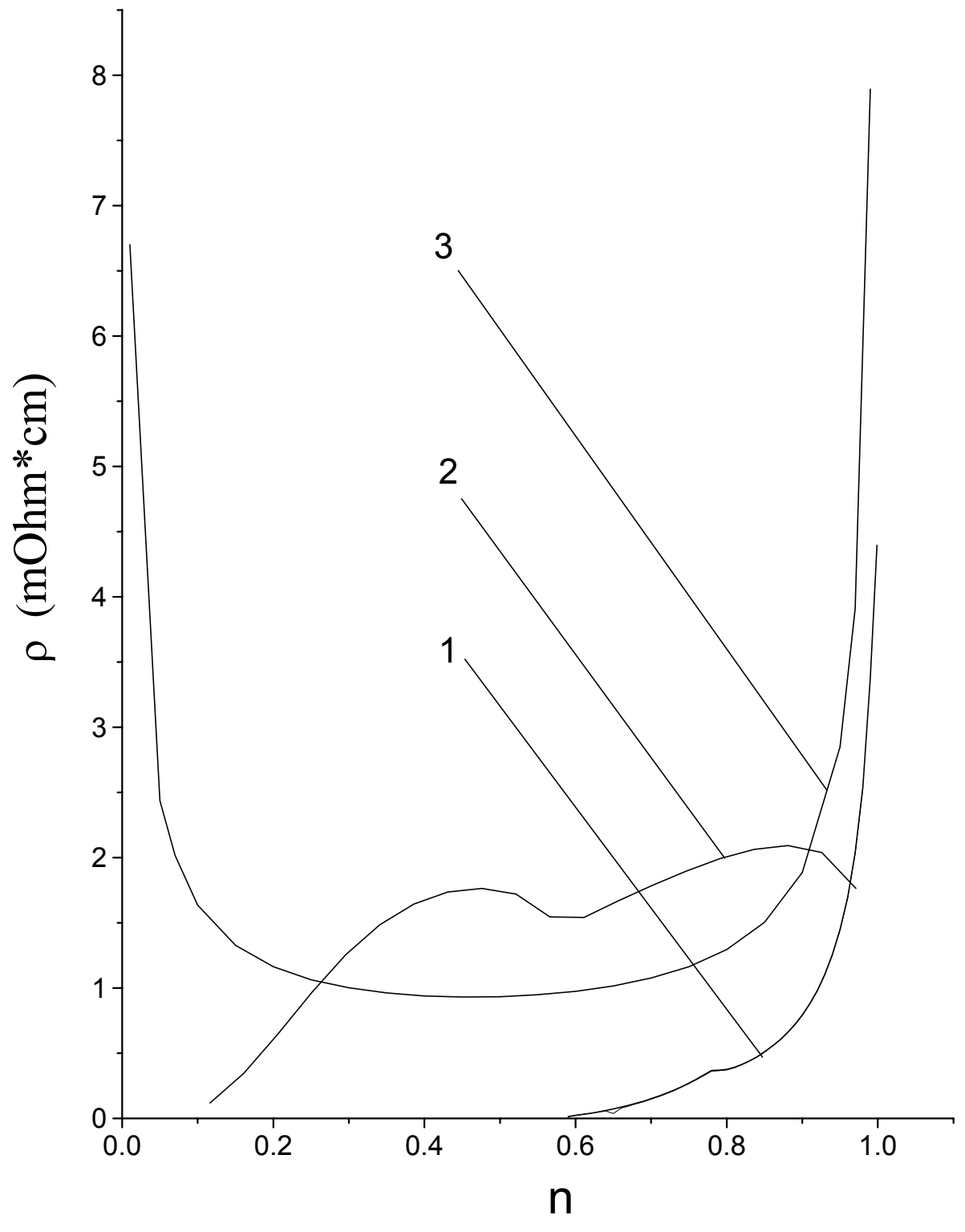

Fig. 11. Resistivity ρ(n) in DE model for FM and PM-2 phases (curves 1 and 2, respectively) at T = 0, h = 0. The curve 3 is obtained using the coherent potential approximation (CPA) [5].

Fig. 12a shows the temperature dependence of resistivity at n = 0.8 neglecting electron-phonon coupling. Unlike the CPA method, a sufficiently abrupt growth of resistivity near the phase transition with increasing T is seen. In Fig. 12b the temperature dependences of imaginary parts $\Lambda_4^{\sigma}(\omega)$ and $\Delta_6^{\sigma}(\omega)$ at $\omega = 0$ are shown. A sharp increase of some dependencies in value near $T_C$ reflects the essential strengthening of scattering processes near the phase transition. In the PM-2 phase the resistivity is weakly dependent on temperature, and its maximal value does not exceed several mOhm*cm as observed in [5]. The authors of Refs. 6,10 and 20 pointed out that it is the characteristic property of systems with dominant DE as for $La_{1-x}Sr_xMnO_3$. In $La_{1-x}Ca_xMnO_3$ compounds the polaron dynamics becomes important because of stronger electron-phonon coupling.

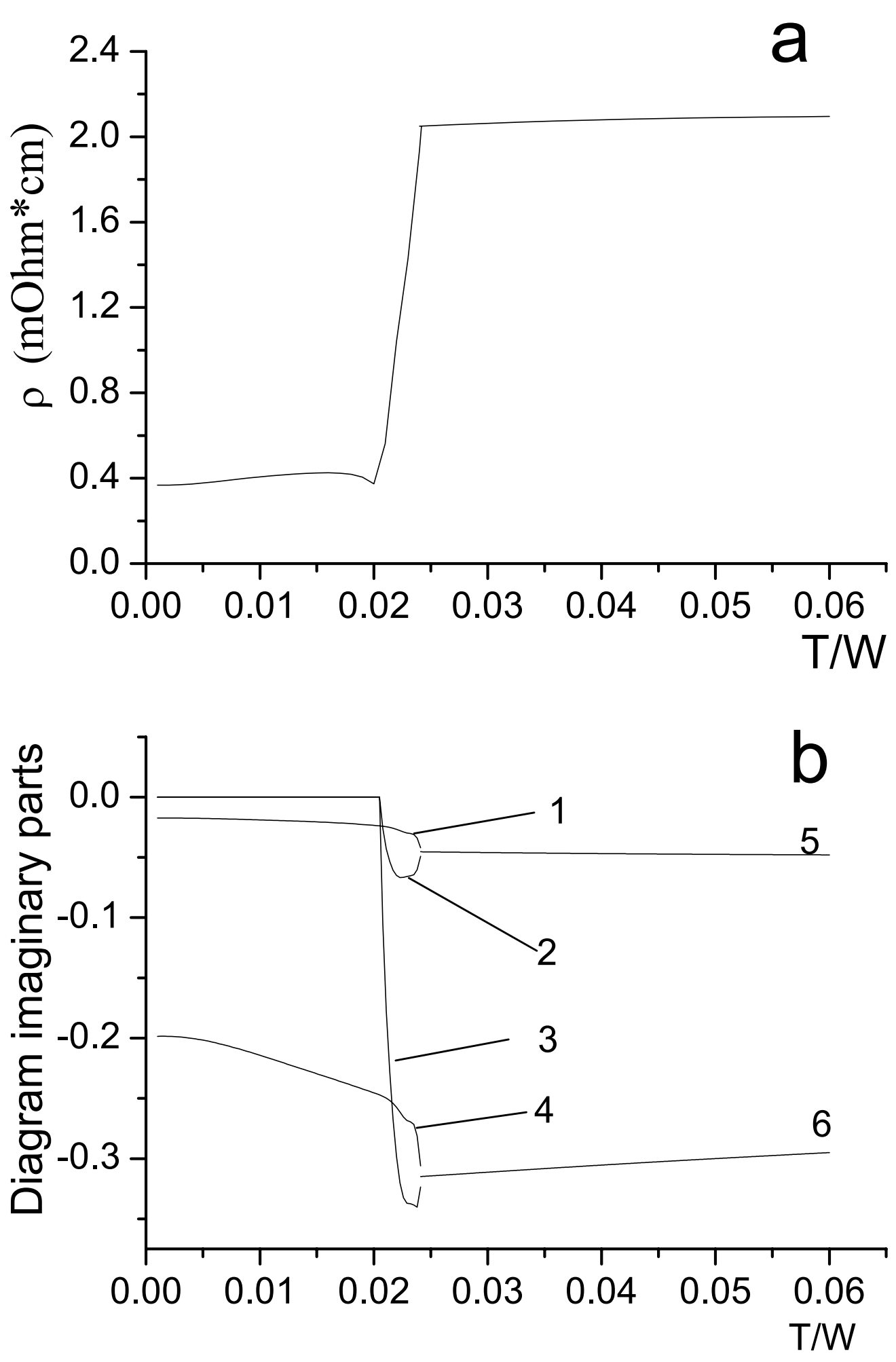

Fig. 12. In DE model the ρ(T) dependences (a) and imaginary parts of $\Delta_6^\sigma(\omega)$ and $\Lambda_4^\sigma(\omega)$ (b) diagrams at n = 0.8 and ω = 0 (curves 1, 3 and 2, 4 for up- and down-spin band, respectively). Curves 5 and 6 are the same plots for $\Delta_6^\sigma(\omega)$ and $\Lambda_4^\sigma(\omega)$ in PM-2 phase, respectively.

In Fig. 13 the influence of weak electron-phonon interaction on ρ(T) dependence calculated using of the first approximation (31) for effective line at n = 0.8 is shown. Indeed, above $T_C$ a large polarons take part in conductivity increasing it with increasing temperature. However, the resistivity falls off with increasing polaron binding energy ξ. A discrepancy between calculated and experimental values of maximum resistivity is due to the influence of vertex corrections in current correlators which are ignored in the bubble approximation.

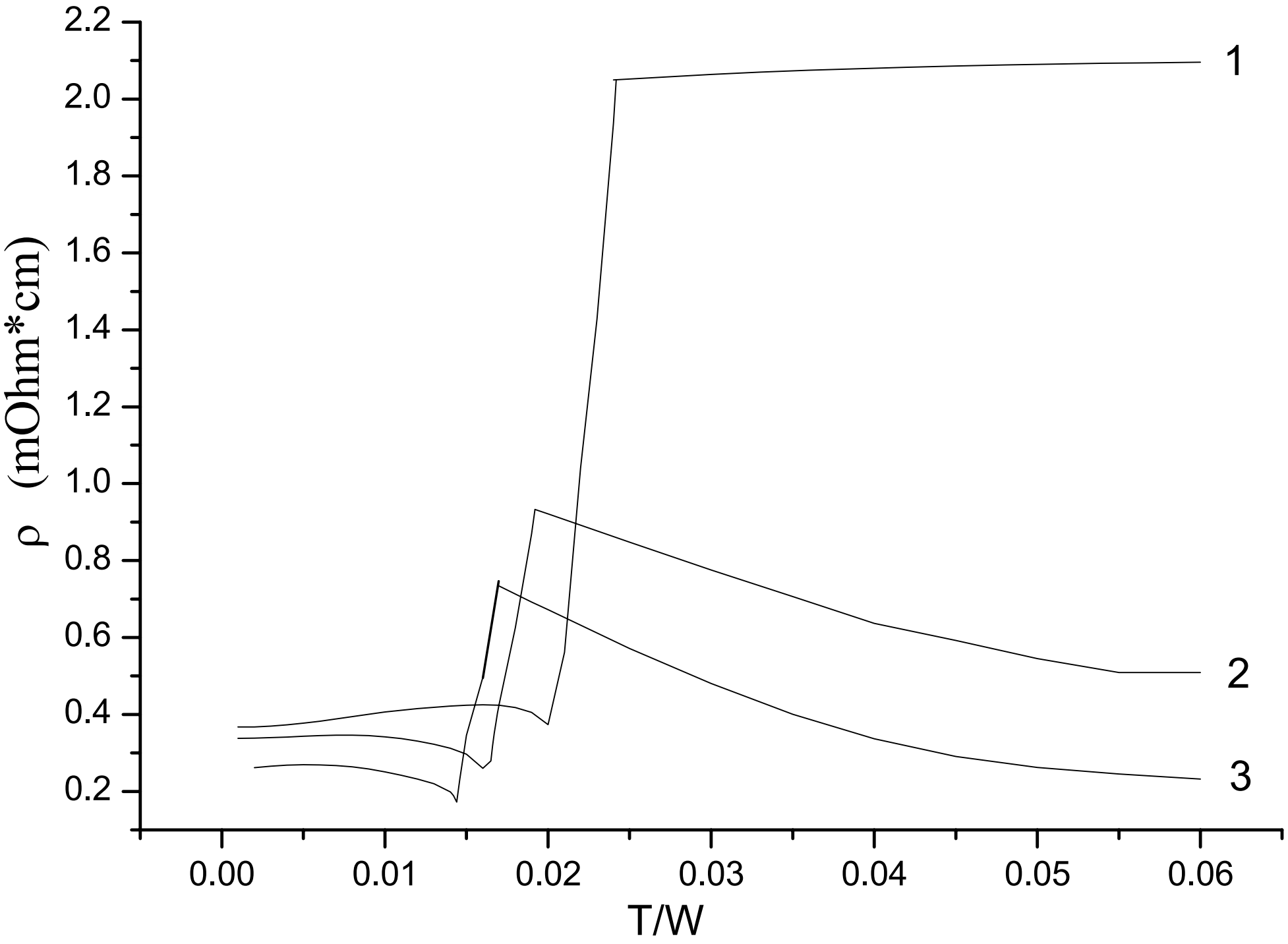

Fig. 13. Influence of weak electron-phonon interaction on ρ(T) at n = 0.8 and h = 0. for polaron binding energy ξ/W : 0, 0.003 and 0.005 (curves 1-3, respectively).

Let us use the second approximation (Eqs. (34)-(35)) to find the contribution of $\Delta_5^\sigma$ diagram to spectral density $R_\sigma(\omega,\mathbf{k})$ calculated using Eq.(51) and taking into account that in this approximation the contribution of $\Lambda_4^\sigma(\omega)$ diagram is equal to zero. It is seen from Fig. 14 that ρ(T) correlates well in value with the experimental maximum resistivity near the phase transition as in the case of $La_{1-x}Ca_xMnO_3$. At high temperatures the chemical potential abruptly decreases that gives zero spectral density. It points to limited usefulness of this approximation in calculation of resitivity.

In Fig. 14 the ρ(T) curves for different magnetic fields, h, at n = 0.8 and ξ/W = 0.2 are shown. The values of both h and superexchange, J(0), were chosen as in Fig.7. In Fig. 14 one can see that the resistivity peak decreases and shifts towards higher fields with increasing applied magnetic field. This agrees with experimental data [23] for the same values of h.

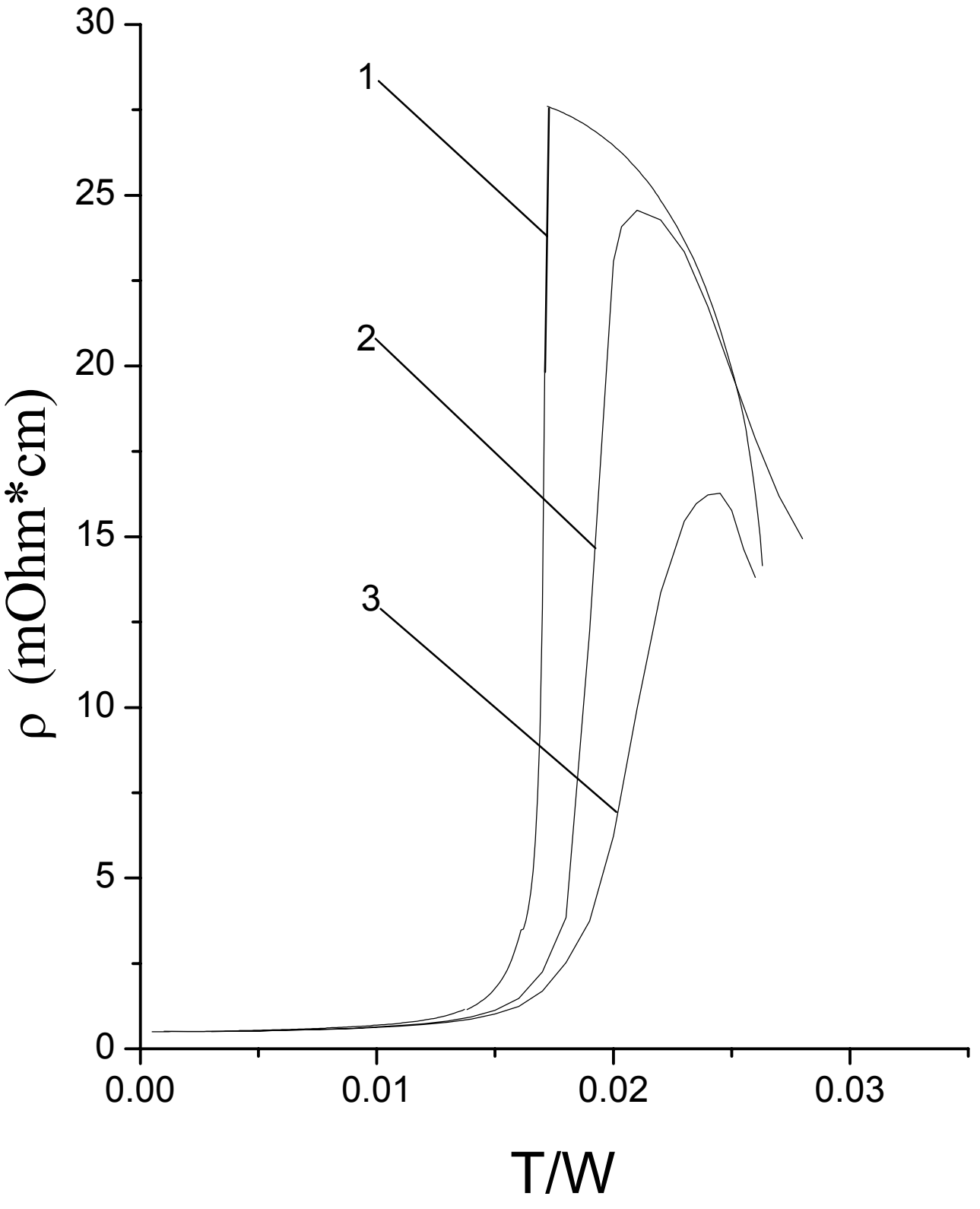

Fig. 14. Temperature dependence of resistivity obtained using the second approximation (Eq. (35)) with account of electron-phonon interaction at $\xi/W = 0.2$, $n = 0.8$ and $h = 0$, $J(0) = 0$ (curve 1) and at $J(0)/W = 0.00215$, $h/W = 0.0001$ and 0.0006 (curves 2 and 3, respectively). Spectral density was calculated using Eq.(51).

It should be noted that the values of h for the same reduction of ρ obtained in works[6,10,20,24] are five times greater than in experiment. The shift of spectral density $R_\sigma(\omega,\mathbf{k})$ peak (Eq.(53)) to electron band edge versus $t(\mathbf{k})$ is the main reason of sharp increase of resistivity. In this case the $\Phi_\sigma(t(\mathbf{k}))t(\mathbf{k})$ function (Eq.(54)) becomes almost antisymmetric.

## 7. Conclusions.

The diagram techniques was constructed taking into account the quantum fluctuations of electron and ion spins in the model of narrow-band Hubbard's magnet with strong Hund's rule coupling. On the basis of separated self-consistent field in DE model with electron-phonon coupling the equations for both magnetization and chemical potential were obtained. The quantum fluctuations of electron and ion spins were shown to reduce the Curie temperature by approximately 30%. In the pure DE model the obtained concentration dependence of $T_C$ in units of bandwidth (W) well coincides with similar result in the CPA method. The analysis of temperature dependence of conductivity in the framework of considered model shows that ρ(T) dependence rapidly drops with decreasing temperature below $T_C$ as in $La_{1-x}Sr_xMnO_3$ with optimal Sr content. This is due to the enforcing of scattering effects near the phase transition. In CPA theory the minimal value of $\rho_0 \sim 1$ mOhm*cm was obtained for PM phase at T = 0. The residual resistivity $\rho_0 << 1$ mOhm*cm was found experimentally for some perovskite manganites. Our calculations agree well with this experimental result near the critical band filling with $n_0 \sim 0.116$ and 0.588 for PM-2 и FM phases, respectively. The influence of electron-phonon coupling on the magnetic structure, ordering temperature, spectral and transport properties using two approximations for effective line of interaction was estimated. Using the first approximation having regard to a few polaron bands with quasigap near the Fermi level, the strong influence of electron-phonon interaction on $T_C$ as well as on magnetization was established. Analysis of spectral density shows the presence of pronounced polaron peaks in the PM-2 phase responsible for increase of conductivity above Tc with increasing temperature. The resistivity falls off with increasing polaron binding energy, ξ. Such an unusual behaviour of ρ(T) is due to miscalculation neglecting the influence of vertex corrections to the two-particle Green's function.

Using the second approximation the effective line of two polaron bands with a gap of the order of polaron binding energy was accounted. The influence of the ξ parameter on both $T_C$ and magnetization was found to be weaker essentially then it is observed in CPA method. This approximation describes well the temperature dependence of resistiviy near the FM-PM phase transition as well as the influence of applied magnetic field on magnetoresistive effect.

**Appendix.**

The annihilation operators are expressed as

$$\begin{aligned} c_{\uparrow} &= L^{9,1} + \frac{\sqrt{3}}{2} L^{10,2} + \frac{1}{\sqrt{2}} L^{11,3} + \frac{1}{2} L^{12,4} \\ c_{\downarrow} &= \frac{1}{2} L^{9,2} + \frac{1}{\sqrt{2}} L^{10,3} + \frac{\sqrt{3}}{2} L^{11,4} + L^{12,5} \end{aligned} \qquad \text{(A.1)}$$

To obtain similar formulae for electron creation operators the Hermitian conjugate of (A.1) was performed. Since we used the wave functions of model space in which the $|\varphi_6\rangle, |\varphi_7\rangle$ and $|\varphi_8\rangle$ functions of ortogonal add-ins are excluded, the equality $c^+c + cc^+ = 1$ fails for Fermi operators. However, it is not important under this consideration because the Wick's theorem is used directly to Hubbard's operators for which the Fermi or Bose origin does not break down.

The operators of electron $n$ and hole $p$ numbers have the form:

$$\begin{aligned} \hat{n} &= L^{1,1} + L^{2,2} + L^{3,3} + L^{4,4} + L^{5,5} \\ \hat{p} &= L^{9,9} + L^{10,10} + L^{11,11} + L^{12,12} \end{aligned} \qquad \text{(A.2)}$$

In this case $p + n = 1$. The operators of electron $n_\sigma$ and hole $p_\sigma$ numbers with spin σ are determined as follows:

$$\hat{n}_{\uparrow} = L^{1,1} + \frac{3}{4}L^{2,2} + \frac{1}{2}L^{3,3} + \frac{1}{4}L^{4,4}$$
$$\hat{n}_{\downarrow} = \frac{1}{4}L^{2,2} + \frac{1}{2}L^{3,3} + \frac{3}{4}L^{4,4} + L^{5,5}$$
$$\hat{p}_{\uparrow} = L^{9,9} + \frac{3}{4}L^{10,10} + \frac{1}{2}L^{11,11} + \frac{1}{4}L^{12,12} \qquad \text{(A.3)}$$
$$\hat{p}_{\downarrow} = \frac{1}{4}L^{9,9} + \frac{1}{2}L^{10,10} + \frac{3}{4}L^{11,11} + L^{12,12}$$

and the next relations

$$\hat{n}_{\uparrow} + \hat{n}_{\downarrow} = \hat{n}\,,\ \hat{p}_{\uparrow} + \hat{p}_{\downarrow} = \frac{5}{4}(1-\hat{n})$$

$$\hat{n}_{\uparrow} - \hat{n}_{\downarrow} = 2\sigma^{z}$$
$$\hat{p}_{\uparrow} - \hat{p}_{\downarrow} = \frac{1}{2}S_0^{z} \qquad \text{(A.4)}$$

take place. The z-projection of spin operator $S^z$ in truncated basis of $t_{2g}$ and $e_g$ electrons is expressed as

$$S^z = \frac{3}{2}(L^{1,1} - L^{5,5} + L^{9,9} - L^{12,12}) + \frac{3}{4}(L^{2,2} - L^{4,4}) + \frac{1}{2}(L^{10,10} - L^{11,11})$$

The imaginary parts of $\Lambda_4^{\sigma}(\omega)$ and $\Delta_6^{\sigma}(\omega)$ functions after analytic continuation to the real axis $i\omega_n \to \omega + i\delta$ are written in the form:

a) approximation (31) with formulae (47) for $\Lambda_4^{\sigma}(\omega)$ and $\Delta_6^{\sigma}(\omega)$:

$$\mathrm{Im}(\Lambda_4^{\sigma}(\omega)) =$$
$$-\frac{3\pi\sigma b(\varepsilon_{-\sigma\sigma}) < \sigma^z + \frac{1}{4}S^z >_0}{W < F^{-\sigma 0} >^2} \sum_{m=-\infty}^{\infty} (\Omega_m(\omega) + \varepsilon_{0-\sigma}) \frac{P_0\varphi_{00}}{P_{-\sigma}\varphi_{m-\sigma}} D_C\left(\frac{6(\Omega_m(\omega) + \varepsilon_{0-\sigma})}{< F^{-\sigma 0} > P_{-\sigma}\varphi_{m-\sigma} W}\right), \qquad \text{(A.5)}$$

where $\Omega_m(\omega) = \omega + \varepsilon_{0-\sigma} + m\omega_0$.

$$\frac{1}{\beta}\mathrm{Im}\left(\Delta_6^{\sigma}(\omega)\right) = -\frac{6\pi\, m(\tilde{\mu})}{W < F^{\sigma 0} >^2} \sum_{m=-\infty}^{\infty} (\Omega_m(\omega) + \varepsilon_{0\sigma})^2 \frac{1}{\left(P_{\sigma}\varphi_{m\sigma}\right)^2} D_C\left(\frac{6(\Omega_m(\omega) + \varepsilon_{0\sigma})}{< F^{-\sigma 0} > P_{\sigma}\varphi_{m\sigma}}\right)$$

(A.6)

б) approximation (34) – (35) with formulae (48) for $\Delta_6^\sigma(\Omega)$ :

$$\frac{1}{\beta}\mathrm{Im}\left(\Delta_6^\sigma(\omega)\right) = -\frac{6\pi m(\tilde{\mu})\left[(\omega+\varepsilon_{0\sigma})^2-\xi^2\right]^2}{W < F^{\sigma 0} >^2 (\omega+\varepsilon_{0\sigma})^2} D_C\left(6\frac{(\omega+\varepsilon_{0\sigma})^2-\xi^2}{W < F^{\sigma 0} > (\omega+\varepsilon_{0\sigma})}\right) \qquad \text{(A.7)}$$

**References.**

**Acknowledgement.**

This work was in part supported by the Polish Government Agency KBN (Project 1 P03B 025 26).